\documentclass[final]{IEEEtran}
\usepackage{amsthm,amssymb,graphicx,multirow,amsmath,color,amsfonts}
\usepackage[update,prepend]{epstopdf}
\usepackage[noadjust]{cite}
\usepackage[latin1]{inputenc}
\usepackage{tikz}
\usepackage{bbm} 
\usepackage[nolist]{acronym}
\usepackage{pdfpages}
\usepackage{multirow}
\usepackage{subfig}
\usepackage{comment}
\def\nb0{{\mathbf{0}}}
\def\nb1{{\mathbf{1}}}

\newtheorem{lemma}{Lemma}

\newtheorem{definition}{Definition}

\newtheorem{theorem}{Theorem}
\newtheorem{prop}{Proposition}

\allowdisplaybreaks 

\begin{acronym}

\acro{5G-NR}{5G New Radio}
\acro{3GPP}{3rd Generation Partnership Project}
\acro{AC}{address coding}
\acro{ACF}{autocorrelation function}
\acro{ACR}{autocorrelation receiver}
\acro{ADC}{analog-to-digital converter}
\acrodef{aic}[AIC]{Analog-to-Information Converter}     
\acro{AIC}[AIC]{Akaike information criterion}
\acro{aric}[ARIC]{asymmetric restricted isometry constant}
\acro{arip}[ARIP]{asymmetric restricted isometry property}
\acro{AoA}{angle of arrival}
\acro{AoD}{angle of departure}
\acro{ARQ}{automatic repeat request}
\acro{AUB}{asymptotic union bound}
\acrodef{awgn}[AWGN]{Additive White Gaussian Noise}     
\acro{AWGN}{additive white Gaussian noise}

\acro{APSK}[PSK]{asymmetric PSK} 

\acro{waric}[AWRICs]{asymmetric weak restricted isometry constants}
\acro{warip}[AWRIP]{asymmetric weak restricted isometry property}
\acro{BCH}{Bose, Chaudhuri, and Hocquenghem}        
\acro{BCHC}[BCHSC]{BCH based source coding}
\acro{BEP}{bit error probability}
\acro{BFC}{block fading channel}
\acro{BG}[BG]{Bernoulli-Gaussian}
\acro{BGG}{Bernoulli-Generalized Gaussian}
\acro{BPAM}{binary pulse amplitude modulation}
\acro{BPDN}{Basis Pursuit Denoising}
\acro{BPPM}{binary pulse position modulation}
\acro{BPSK}{binary phase shift keying}
\acro{BPZF}{bandpass zonal filter}
\acro{BSC}{binary symmetric channels}              
\acro{BU}[BU]{Bernoulli-uniform}
\acro{BER}{bit error rate}
\acro{BS}{base station}

\acro{CP}{Cyclic Prefix}
\acrodef{cdf}[CDF]{cumulative distribution function}   
\acro{CDF}{cumulative distribution function}
\acrodef{c.d.f.}[CDF]{cumulative distribution function}
\acro{CCDF}{complementary cumulative distribution function}
\acrodef{ccdf}[CCDF]{complementary CDF}               
\acrodef{c.c.d.f.}[CCDF]{complementary cumulative distribution function}
\acro{CD}{compliance distance}

\acro{CDMA}{Code Division Multiple Access}
\acro{ch.f.}{characteristic function}
\acro{CIR}{channel impulse response}
\acro{cosamp}[CoSaMP]{compressive sampling matching pursuit}
\acro{CR}{cognitive radio}
\acro{cs}[CS]{compressed sensing}                   
\acrodef{cscapital}[CS]{Compressed sensing}
\acrodef{CS}[CS]{compressed sensing}
\acro{CSI}{channel state information}

\acro{CCSDS}{consultative committee for space data systems}
\acro{CC}{convolutional coding}

\acro{DAA}{detect and avoid}
\acro{DAB}{digital audio broadcasting}
\acro{DCT}{discrete cosine transform}
\acro{dft}[DFT]{discrete Fourier transform}
\acro{DR}{distortion-rate}
\acro{DS}{direct sequence}
\acro{DS-SS}{direct-sequence spread-spectrum}
\acro{DTR}{differential transmitted-reference}
\acro{DVB-H}{digital video broadcasting\,--\,handheld}
\acro{DVB-T}{digital video broadcasting\,--\,terrestrial}
\acro{DL}{downlink}
\acro{DSSS}{Direct Sequence Spread Spectrum}
\acro{DFT-s-OFDM}{Discrete Fourier Transform-spread-Orthogonal Frequency Division Multiplexing}
\acro{DAS}{distributed antenna system}
\acro{DNA}{Deoxyribonucleic Acid}

\acro{EC}{European Commission}
\acro{EED}[EED]{exact eigenvalues distribution}
\acro{EIRP}{Equivalent Isotropically Radiated Power}
\acro{ELP}{equivalent low-pass}
\acro{eMBB}{Enhanced Mobile Broadband}
\acro{EMFE}{electromagnetic field exposure}
\acro{EU}{European union}

\acro{FC}[FC]{fusion center}
\acro{FCC}{Federal Communications Commission}
\acro{FEC}{forward error correction}
\acro{FFT}{fast Fourier transform}
\acro{FH}{frequency-hopping}
\acro{FH-SS}{frequency-hopping spread-spectrum}
\acrodef{FS}{Frame synchronization}
\acro{FSsmall}[FS]{frame synchronization}  
\acro{FDMA}{Frequency Division Multiple Access}

\acro{GA}{Gaussian approximation}
\acro{GF}{Galois field }
\acro{GG}{Generalized-Gaussian}
\acro{GIC}[GIC]{generalized information criterion}
\acro{GLRT}{generalized likelihood ratio test}
\acro{GPS}{Global Positioning System}
\acro{GMSK}{Gaussian minimum shift keying}
\acro{GSMA}{Global System for Mobile communications Association}

\acro{HAP}{high altitude platform}

\acro{IDR}{information distortion-rate}
\acro{IFFT}{inverse fast Fourier transform}
\acro{iht}[IHT]{iterative hard thresholding}
\acro{i.i.d.}{independent, identically distributed}
\acro{IoT}{Internet of Things}                      
\acro{IR}{impulse radio}
\acro{lric}[LRIC]{lower restricted isometry constant}
\acro{lrict}[LRICt]{lower restricted isometry constant threshold}
\acro{ISI}{intersymbol interference}
\acro{ITU}{International Telecommunication Union}
\acro{ICNIRP}{International Commission on Non-Ionizing Radiation Protection}
\acro{IEEE}{Institute of Electrical and Electronics Engineers}
\acro{ICES}{IEEE international committee on electromagnetic safety}
\acro{IEC}{international Electrotechnical Commission}
\acro{IARC}{International Agency on Research on Cancer}
\acro{IS-95}{Interim Standard 95}

\acro{KPI}{Key Performance Indicator}

\acro{LEO}{low earth orbit}
\acro{LF}{likelihood function}
\acro{LLF}{log-likelihood function}
\acro{LLR}{log-likelihood ratio}
\acro{LLRT}{log-likelihood ratio test}
\acro{LoS}{line-of-sight}
\acro{NLoS}{non-line-of-sight}
\acro{LRT}{likelihood ratio test}
\acro{wlric}[LWRIC]{lower weak restricted isometry constant}
\acro{wlrict}[LWRICt]{LWRIC threshold}
\acro{LPWAN}{low power wide area network}
\acro{LoRaWAN}{Low power long Range Wide Area Network}

\acro{MB}{multiband}
\acro{MC}{multicarrier}
\acro{MDS}{mixed distributed source}
\acro{MF}{matched filter}
\acro{m.g.f.}{moment generating function}
\acro{MI}{mutual information}
\acro{MIMO}{multiple-input multiple-output}
\acro{MISO}{multiple-input single-output}
\acrodef{maxs}[MJSO]{maximum joint support cardinality}                       
\acro{ML}[ML]{maximum likelihood}
\acro{MMSE}{minimum mean-square error}
\acro{MMV}{multiple measurement vectors}
\acrodef{MOS}{model order selection}
\acro{M-PSK}[${M}$-PSK]{$M$-ary phase shift keying}                       
\acro{M-APSK}[${M}$-PSK]{$M$-ary asymmetric PSK} 

\acro{M-QAM}[$M$-QAM]{$M$-ary quadrature amplitude modulation}
\acro{MRC}{maximal ratio combiner}                  
\acro{maxs}[MSO]{maximum sparsity order}                                      
\acro{M2M}{machine to machine}                                                
\acro{MUI}{multi-user interference}
\acro{mMTC}{massive Machine Type Communications}      
\acro{mm-Wave}[mm-Wave]{millimeter-wave}
\acro{MP}{mobile phone}
\acro{MPE}{maximum permissible exposure}
\acro{MAC}{media access control}
\acro{NB}{narrowband}
\acro{NBI}{narrowband interference}
\acro{NLA}{nonlinear sparse approximation}
\acro{NLOS}{Non-Line of Sight}
\acro{NTIA}{National Telecommunications and Information Administration}
\acro{NTP}{National Toxicology Program}

\acro{OC}{optimum combining}                             
\acro{OC}{optimum combining}
\acro{ODE}{operational distortion-energy}
\acro{ODR}{operational distortion-rate}
\acro{OFDM}{orthogonal frequency-division multiplexing}
\acro{omp}[OMP]{orthogonal matching pursuit}
\acro{OSMP}[OSMP]{orthogonal subspace matching pursuit}
\acro{OQAM}{offset quadrature amplitude modulation}
\acro{OQPSK}{offset QPSK}
\acro{OFDMA}{Orthogonal Frequency-division Multiple Access}

\acro{OQPSK/PM}{OQPSK with phase modulation}

\acro{PAM}{pulse amplitude modulation}
\acro{PAR}{peak-to-average ratio}
\acrodef{pdf}[PDF]{probability density function}                      
\acro{PDF}{probability density function}
\acrodef{p.d.f.}[PDF]{probability distribution function}
\acro{PDP}{power dispersion profile}
\acro{PMF}{probability mass function}                             
\acrodef{p.m.f.}[PMF]{probability mass function}
\acro{PN}{pseudo-noise}
\acro{PPM}{pulse position modulation}
\acro{PRake}{Partial Rake}
\acro{PSD}{power spectral density}
\acro{PSEP}{pairwise synchronization error probability}
\acro{PSK}{phase shift keying}
\acro{PD}{power density}
\acro{8-PSK}[$8$-PSK]{$8$-phase shift keying}
\acro{PHP}{Poisson hole process}
 \acro{PPP}{Poisson point process}
\acrodefplural{PPP}{Poisson point processes}
\acro{FSK}{frequency shift keying}

\acro{QAM}{Quadrature Amplitude Modulation}
\acro{QPSK}{quadrature phase shift keying}
\acro{OQPSK/PM}{OQPSK with phase modulator }

\acro{RD}[RD]{raw data}
\acro{RDL}{"random data limit"}
\acro{ric}[RIC]{restricted isometry constant}
\acro{rict}[RICt]{restricted isometry constant threshold}
\acro{rip}[RIP]{restricted isometry property}
\acro{ROC}{receiver operating characteristic}
\acro{rq}[RQ]{Raleigh quotient}
\acro{RS}[RS]{Reed-Solomon}
\acro{RSC}[RSSC]{RS based source coding}
\acro{r.v.}{random variable}                               
\acro{R.V.}{random vector}
\acro{RMS}{root mean square}
\acro{RFR}{radiofrequency radiation}
\acro{RIS}{reconfigurable intelligent surface}

\acro{SA}[SA-Music]{subspace-augmented MUSIC with OSMP}
\acro{SCBSES}[SCBSES]{Source Compression Based Syndrome Encoding Scheme}
\acro{SCM}{sample covariance matrix}
\acro{SEP}{symbol error probability}
\acro{SIMO}{single-input multiple-output}
\acro{SINR}{signal-to-interference-plus-noise ratio}
\acro{SIR}{signal-to-interference ratio}
\acro{SISO}{single-input single-output}
\acro{SMV}{single measurement vector}
\acro{SNR}[\textrm{SNR}]{signal-to-noise ratio} 
\acro{sp}[SP]{subspace pursuit}
\acro{SS}{spread spectrum}
\acro{SW}{sync word}
\acro{SAR}{specific absorption rate}
\acro{SSB}{synchronization signal block}
\acro{SDG}{Sustainable Development Goal}
\acro{SBF}{spatial bandpass filtering}
\acro{SG}{stochastic geometry}

\acro{TH}{time-hopping}
\acro{ToA}{time-of-arrival}
\acro{TR}{transmitted-reference}
\acro{TW}{Tracy-Widom}
\acro{TWDT}{TW Distribution Tail}
\acro{TCM}{trellis coded modulation}
\acro{TDD}{time-division duplexing}
\acro{TDMA}{Time Division Multiple Access}

\acro{UAV}{unmanned aerial vehicle}
\acro{uric}[URIC]{upper restricted isometry constant}
\acro{urict}[URICt]{upper restricted isometry constant threshold}
\acro{UWB}{ultrawide band}
\acro{UWBcap}[UWB]{Ultrawide band}   
\acro{URLLC}{Ultra Reliable Low Latency Communications}
         
\acro{wuric}[UWRIC]{upper weak restricted isometry constant}
\acro{wurict}[UWRICt]{UWRIC threshold}                
\acro{UE}{user equipment}
\acro{UL}{uplink}

\acro{ULA}{uniform linear array}
\acro{UPA}{uniform planar array}

\acro{WiM}[WiM]{weigh-in-motion}
\acro{WLAN}{wireless local area network}

\acro{wm}[WM]{Wishart matrix}                               
\acroplural{wm}[WM]{Wishart matrices}
\acro{WMAN}{wireless metropolitan area network}
\acro{WPAN}{wireless personal area network}
\acro{wric}[WRIC]{weak restricted isometry constant}
\acro{wrict}[WRICt]{weak restricted isometry constant thresholds}
\acro{wrip}[WRIP]{weak restricted isometry property}
\acro{WSN}{wireless sensor network}                        
\acro{WSS}{wide-sense stationary}
\acro{WHO}{World Health Organization}
\acro{WP}{work package}

\acro{sss}[SpaSoSEnc]{sparse source syndrome encoding}
\acro{SO}{strategic objective}

\acro{VLC}{visible light communication}
\acro{RF}{radio frequency}
\acro{FSO}{free space optics}
\acro{IoST}{Internet of space things}

\acro{GSM}{Global System for Mobile Communications}
\acro{2G}{second-generation cellular network}
\acro{3G}{third-generation cellular network}
\acro{4G}{fourth-generation cellular network}
\acro{5G}{5th-generation cellular network}	
\acro{gNB}{next generation node B base station}
\acro{NR}{New Radio}
\acro{UN}{United Nations}
\acro{UMTS}{Universal Mobile Telecommunications Service}
\acro{LTE}{Long Term Evolution}

\acro{QoS}{Quality of Service}
\end{acronym}

\begin{document}
\graphicspath{{./figures/}}

\title{
Joint Coverage and Electromagnetic Field Exposure Analysis in Downlink and Uplink for RIS-Assisted Networks  
}
\author{Lin Chen,~\IEEEmembership{Graduate Student Member,~IEEE}, Ahmed~Elzanaty,~\IEEEmembership{Senior Member,~IEEE}, Mustafa~A.~Kishk,~\IEEEmembership{Member,~IEEE}, and Ying-Jun Angela Zhang,~\IEEEmembership{Fellow,~IEEE} 
\thanks{This work was supported in part by the General Research Fund (project number 14202421, 14214122, 14202723, 14207624), Area of Excellence Scheme grant (project number AoE/E-601/22-R), and NSFC/RGC Collaborative Research Scheme (project number CRS\_HKUST603/22, CRS\_HKU702/24), all from the Research Grants Council of Hong Kong.
Part of this work was supported by 6G-FINESSE project under the CHEDDAR Hub.
The views expressed are those of the authors and do not necessarily represent
the project.
An earlier version of this paper
was presented in part at the IEEE Wireless Communications and Networking
Conference (WCNC), 2024, Dubai, United Arab Emirates~\cite{wcnc}. 
}
\thanks{Lin Chen and Ying-Jun Angela Zhang are with the Department of Information Engineering, The Chinese University of Hong Kong (CUHK), Hong Kong (e-mail: \{cl022,yjzhang\}@ie.cuhk.edu.hk). }
\thanks{A. Elzanaty is with the 5GIC \& 6GIC, Institute for Communication Systems (ICS), University of Surrey, Guildford, GU2 7XH, United Kingdom (email: a.elzanaty@surrey.ac.uk).}
\thanks{M. A. Kishk is with the Department of Electronic Engineering, Maynooth University, Maynooth, W23 F2H6, Ireland (email: mustafa.kishk@mu.ie).}
}

\maketitle

\begin{abstract}
\Acp{RIS} have shown the potential to improve \ac{SINR} related coverage, especially at high-frequency communications.
However, assessing \ac{EMFE} and establishing \ac{EMFE} regulations in RIS-assisted large-scale networks remain open issues. 
{This paper proposes a \ac{SG} based framework to characterize \ac{SINR} and \ac{EMFE} in such networks for downlink and uplink scenarios.}
Particularly, we carefully consider the association rule with the presence of RISs, accurate antenna pattern at base stations (BSs), fading model, and power control mechanism at mobile devices in the system model.
Under the proposed framework, we derive the marginal and joint distributions of SINR and \ac{EMFE} in downlink and uplink, respectively.  
The first moment of \ac{EMFE} is also provided. Additionally, we design the \ac{CD} between a BS/RIS and a user to comply with the \ac{EMFE} regulations. To facilitate efficient identification, we further provide approximate closed-form expressions for \acp{CD}.
From numerical results of the marginal distributions, we find that in the downlink scenario, deploying RISs may not always be beneficial, as the improved SINR comes at the cost of increased \ac{EMFE}. However, in the uplink scenario, RIS deployment is promising to enhance coverage while still maintaining \ac{EMFE} compliance.
By simultaneously evaluating coverage and compliance metrics through joint distributions, we demonstrate the feasibility of RISs in improving uplink and downlink performance.
Insights from this framework can contribute to establishing \ac{EMFE} guidelines and achieving a balance between coverage and compliance when deploying RISs.

\end{abstract}

\begin{IEEEkeywords}
\ac{EMFE}, \ac{SINR}, coverage, compliance distance, RIS, joint distribution, stochastic geometry, millimeter wave.
\end{IEEEkeywords}

\acresetall
\section{Introduction}\label{sec:intro}

Radio frequency (RF) radiation emitted by \acp{BS} (in downlink) and mobile devices (in uplink) generates \ac{EMFE} in cellular networks.
If not controlled, \ac{EMFE} can have potential adverse thermal effects on exposed tissues, endangering population health.
Regulatory authorities (e.g., \ac{ICNIRP}~\cite{ICNIRPGuidelines:18} and \ac{FCC}~\cite{FCCEvaluatingCom:97}) have implemented regulations to ensure \ac{EMFE} within a safe limit in current cellular networks. This includes the establishment of the \ac{CD} between a \ac{BS} and a user.
If a user enters an area centered at a \ac{BS} with a radius of the \ac{CD}, the corresponding \ac{EMFE} may exceed the safe limit, and user safety cannot be guaranteed.
Hence, the assessment of \ac{EMFE} and the establishment of \ac{EMFE} regulations are crucial in developing future cellular networks~\cite{Risk5G}.
 
A novel technology, \ac{RIS}, consisting of numerous passive elements, can reflect incident waves into specific directions by dynamically adjusting the element phases~\cite{tutorialRIS}. This enables RISs to establish a cascaded \ac{LoS} link between a \ac{BS} and a user and provide directional transmission gain (i.e., RIS gain), effectively alleviating blockage effects and compensating for path loss.
Hence, RISs hold significant promise for extending coverage of future cellular networks, especially at high-frequency communications~\cite{Millimeter}.
The potential impact of RISs on \ac{EMFE} has also attracted great attention. Under the constraints of \ac{EMFE}, researchers optimized the BS beamforming and/or \ac{RIS} phases to maximize the \ac{SINR} or capacity in a single cell with a RIS~\cite{R3,R4} or multiple RISs~\cite{R1-2}. 
On the other hand, under the constraints of \ac{SINR}, spectral efficiency, or throughput, researchers designed \ac{RIS} phases to minimize uplink and/or downlink exposure in a single cell with a RIS~\cite{R2,dos2023emf,chemingui2024emf} or multiple RISs~\cite{10345757}.
The authors in \cite{R1-1} extended the analysis to multi-cell scenarios, where BSs and RISs are deployed at specific locations.
These studies highlight the necessity of considering the impact of RISs on \ac{EMFE} and underscore the importance of jointly assessing this impact alongside \ac{SINR}.
However, these studies either focus on small-scale analysis (neglecting \ac{EMFE} and interference from neighboring cells) or fail to adequately characterize the spatial randomness of transmission nodes and obstacles and the spatial characteristics of transmission links.

Comprehensive large-scale analysis on RIS-assisted multi-cell networks with spatial randomness is an unexplored yet critical research area. RISs, with the capability of concentrating the energy of incident waves into a specific direction, introduce potential \ac{EMFE} risks, particularly in that concentrated direction.  
{In this context, several crucial questions arise. \textbf{Q1}: Does deploying large-scale RISs exacerbate \ac{EMFE}? \textbf{Q2}: Should a specific \ac{CD} be established for a RIS? \textbf{Q3}: How do RISs impact network SINR and \ac{EMFE} in downlink/uplink simultaneously?
\textbf{Q4}: What deployment strategy can enhance SINR while effectively mitigating \ac{EMFE}?}
We aim to answer these questions by providing marginal and joint distributions of downlink/uplink \ac{SINR} and \ac{EMFE} in large-scale RIS-assisted networks.
Particularly, the spatial randomness of networks is characterized using tools of \ac{SG}~\cite{SGtutorial,andrews2016primer}.

\subsection{Related Works}\label{subsec:related}

\Ac{SG} is a useful tool for modeling large-scale network topologies with randomly distributed nodes and analyzing performance metrics~\cite{SGtutorial,andrews2016primer}.
This tool has recently been applied to model RIS-assisted cellular networks and analyze the distribution of SINR.
The authors in~\cite{kishk2020exploiting} modeled the locations of \acp{RIS}, \acp{BS}, and users as homogeneous \acp{PPP} and modeled RIS orientations under a line Boolean model. 
Considering that RISs can provide cascaded LoS links when the direct link between a user and its nearest BS is \ac{NLoS}~\cite{RIS_model,distributedRIS,10412176}, the statistics of cascaded links are characterized by the reflection probability of the RIS, the association probability, and the distance distributions.
The authors in~\cite{IEEEAccess,distributedRIS,10412176} derived the \ac{CCDF} of SINR of the networks consisting of BSs with \acp{ULA}  and RISs.  
Under a practical \ac{CSI} assumption that the channel angle information is known, 
a simplified but not accurate antenna model (flat-top
pattern) was adopted to characterize directional transmission gain. The above works have demonstrated the feasibility of RISs in
improving downlink communication quality. 
Additionally, considering the fixed transmit power at mobile devices, RISs can enhance uplink coverage~\cite{9977979}.

In large-scale networks without RISs, researchers have characterized the \ac{CDF} of \ac{EMFE} using tools from \ac{SG} for omnidirectional antennas~\cite{Joint,qin2023unveiling} and directional antennas~\cite{R2-1,muhammad2021stochastic}. 
Under the distance-dependent power control mechanism at the mobile devices, the authors in \cite{Joint,qin2023unveiling} evaluated uplink \ac{EMFE}. Assuming the perfect \ac{CSI} (including not only angle information) for beamforming, the downlink \ac{EMFE} for multi-antenna BSs was analyzed~\cite{R2-1}. 
The authors in~\cite{muhammad2021stochastic} analyzed the \ac{CDF} of downlink \ac{EMFE} (measured by received power density at the typical user) from macro-cell \acp{BS} and small-cell \acp{BS} under a flat-top/sectored antenna patterns based on the channel angle information. Moreover, the BS \ac{CD} (i.e., the minimum allowable separation between a BS and a user which ensures that the corresponding \ac{EMFE} must not exceed the maximum allowable limit with a predefined probability) was identified by solving an optimization problem related to the \ac{CDF} of downlink \ac{EMFE} in~\cite{Joint}. 

To understand the trade-off between increasing SINR and reducing \ac{EMFE}, investigating their joint distribution is crucial. 
Previous research has conducted similar joint analyses of downlink SINR and received power in simultaneous wireless information and power transfer networks, where the directional transmission is characterized by the flat-top pattern for analytical simplicity~\cite{R1-3,R1-4}. Recently, works in \cite{gontier2023joint, R1-5,R2-3} jointly characterized SINR and \ac{EMFE} (received power density) in the downlink and/or uplink of networks with omnidirectional antennas and an uplink power control mechanism. The downlink joint analysis was extended to the scenarios with the multi-antenna BSs under the perfect \ac{CSI} assumption \cite{R2-2}.
These joint analyses are valuable in demonstrating whether deployment strategies aimed at improving SINR could inadvertently exacerbate \ac{EMFE}.

Extending the above \ac{EMFE} analysis to large-scale RIS-assisted networks is not direct and is lacking in the literature.
First, to accurately assess \ac{EMFE}, it is crucial to characterize the spatial statistics of the transmission links, including the traditional direct link and the RIS-involved cascaded link, and their correlations.
Moreover, RISs are often deployed alongside multi-antenna BSs to leverage directional beamforming gains.
This requires precise modeling of the directional antenna pattern to accurately capture the spatial power distribution, a challenge not fully addressed in existing studies.
Another missing but crucial aspect is the setup of the CD between a RIS and a user. This distance needs to be carefully managed to ensure public safety in the practical deployment of RISs. 
Moreover, the impact of RISs on the traditional distance-dependent power control mechanism in the uplink remains unclear.
Compared with the direct link,  the path loss in the cascaded link can be compensated by not only the transmit power but also the RIS gain, making uplink power control dependent on both the distance and RIS gain.
Incorporating this RIS-assisted mechanism into the modeling and analysis is essential but has not been thoroughly explored in the literature.
Furthermore, a joint analysis of SINR and \ac{EMFE} to fully understand the impact of large-scale RIS deployment is also underdeveloped.

\subsection{Contributions}\label{subsec:contri}

Motivated by the above discussions, this paper proposes a general framework for characterizing the marginal and joint distributions of SINR and \ac{EMFE} in RIS-assisted networks and establishing \acp{CD} under \ac{EMFE} regulations, where the precise directional antenna modeling and the refinement of the power control mechanism are incorporated.
Specifically, the random locations of BSs, RISs, and users are modeled \acp{PPP}.
Moreover, based on the channel angle information, each multi-antenna BS adopts analog beamforming to align the beam direction towards its associated user (or RIS) in the direct (or cascaded) link, and the phases of a RIS are designed to reflect the beam from its associated BS towards its associated user.
The main contributions are listed as follows.
\begin{itemize}
\item We provide a general system model of RIS-assisted networks, including an accurate antenna pattern for analog beamforming, a line Boolean model for blockage effects, the Nakagami-m fading model for \ac{LoS} and \ac{NLoS} conditions, and the RIS-involved power control mechanism at mobile devices. 
Based on this model, we provide the spatial
characteristics of transmission links by analyzing the interdependent distance distributions within a triangle formed by the associated BS, RIS, and user.

\item We derive the marginal distributions of downlink SINR and \ac{EMFE}, along with the expectation of \ac{EMFE}. Specifically, to address the complexity introduced by a complicated beam pattern when capturing the spatial signal power distribution, we adopt a discrete antenna pattern for accurate and tractable performance analysis. 
Furthermore, by capturing the correlation between SINR and \ac{EMFE}, e.g., path loss under the interdependent distance distributions and small-scale fading under LoS/NLoS conditions, we derive the joint distribution of downlink SINR and \ac{EMFE}.

\item We extend our analysis to the marginal and joint distributions of uplink SINR and \ac{EMFE}.
Compared with the direct link, the path loss in the cascaded link can be compensated by the RIS gain and the transmit power. 
This RIS-assisted power control mechanism further complicates the characterization of uplink interference since the dynamic transmit power of an interfering user is related to the type and distances of the serving (direct/cascaded) link between the interfering user and its own associated BS.
We propose an approximation method to simplify the analysis while maintaining accuracy.

\item To design the \acp{CD} for RISs, we formulate optimization problems based on the \ac{CDF} of downlink \ac{EMFE}.
Note that the \ac{EMFE} induced by a RIS (intermediate reflective node) depends on the distance to its associated \ac{BS}.
Given the intricate nature of the \ac{CDF} of downlink EMFE, which involves the serving and interfering signal power density distributions and the interdependent distance distribution, deriving a closed-form expression for \acp{CD} is challenging.
To streamline the CD identification, we provide a closed-form approximation by focusing on the dominant \ac{EMFE} caused by serving signals. 
\end{itemize}

The above analytical framework, including the marginal and joint distributions of downlink/uplink SINR and EMFE as well as the CD identification, offers a holistic understanding of how RISs affect coverage and compliance performance, guiding the deployment of SINR-efficient and EMFE-safe RIS-assisted networks.

The remainder of this paper is organized as follows.
In Sec.~\ref{sec:sys}, we introduce the system model.
We provide the modeling of the downlink \ac{SINR} and \ac{EMFE} and define corresponding performance metrics in Sec.~\ref{sec:DLmodel}. We then analyze those metrics in Sec.~\ref{sec:DLanaly}. 
We also present the modeling of the uplink \ac{SINR} and \ac{EMFE} in Sec.~\ref{sec:ULmodel} and provide uplink analysis in Sec.~\ref{sec:ULanaly}. 
{Sec.~\ref {sec:results} discusses the numerical results and provides answers to \textbf{Q1-Q4}.} Finally, Sec.~\ref{sec:conclusion} summarizes the paper. 
A summary of notations is provided in Table~\ref{tab:TableOfNotations}. 
We use the superscript ``$'$'' to distinguish downlink and uplink variables.

\begin{table}[t]\caption{Table of notations.}
\vspace{-3mm}
\centering
\begin{center}
{ \linespread{1}
\renewcommand{\arraystretch}{1.3}
    \begin{tabular}{ {c} | {c} }
    \hline
        \hline
    \textbf{Notation} & \textbf{Description} \\ \hline
    $\Psi_{\rm b}$; $\Psi_{\rm u}$ & PPP modeling the locations of BSs or users. \\ \hline
    $\Psi_{\rm r}$; $\Psi_{\rm o}$ & PPP modeling the locations of RISs or obstacles.\\ \hline
    $\lambda_l$ & The density of $\Psi_{l}$, $l\in\{\rm b,u,r,o\}$.   \\ \hline
    $\mu$ & The fraction of obstacles equipped with RISs.
    \\ \hline
    $N_{\rm b}$; $N_{\rm r}$ & The element number of each BS antenna array or RIS.\\ \hline
    $G_{\rm b}$; $G_{\rm r}$ & Maximum gain provided by BS antenna array or RIS.\\ \hline
    $\Delta$ & Spatial AoD deviation off the aligned direction. \\ \hline
    ${\rm DL,CL,DN}$ & Direct LoS, cascaded LoS, and direct NLoS.\\ \hline
    $q,\mathcal{S} $ & Type of serving links, $q\in \mathcal{S}=\{\rm DL,CL,DN\}$.\\ \hline
    $\mathcal{A}_q$ & Association probability of each type of serving link. \\ \hline
    $v$ & LoS or NLoS, $v\in\{\rm L,N\}$.\\ \hline
    $\epsilon$ & Power control factor. \\ \hline
    $\lambda_f$ & Wavelength ($\lambda_f$) of the center carrier frequency ($f$). \\ \hline
    $\mathcal{E}$ & Antenna effective area $\mathcal{E} = {\lambda_f^2}/{4\pi}$. \\ \hline
    $\zeta$ & Reference path loss $\zeta=({\lambda_f}/{4 \pi })^2$.  \\ \hline
    $\beta$ & Blockage parameter. \\ \hline
    $\alpha_{\rm L}$; $\alpha_{\rm N}$ & Path-loss exponents in LoS or NLoS links.  \\ \hline
    $m_{v}$, $m_{q}$ & Shaping parameter of Nakagami-m fading model.\\ \hline  
    $H_{q}$; $H'_{q}$ & Small-scale fading coefficient in downlink or uplink. \\ \hline
    $\Upsilon$, $\gamma$; $\Upsilon'$, $\gamma'$ & SINR and threshold in downlink or uplink.\\ \hline  
    $\sigma^2$; $\sigma'^2$ & Noise power in downlink or uplink.  \\ \hline
    $p_{\rm b}$; $p_{\rm u}$ & Transmit power of a BS or user.  \\ \hline
    $\mathcal{W}$, $\omega$; $\mathcal{W}'$, $\omega'$ & \ac{EMFE} and constraint in downlink or uplink. \\ \hline
    $\mathcal{W}_1$;  $\mathcal{W}_2$ & Downlink  exposure from serving or interfering signals. \\ \hline
    $\bar F_T$; $F_T$; $f_{T}$ & CCDF of $T$; CDF of $T$; PDF of $T$. \\ \hline
    $F^{-1}_T$ & Inverse function of $F_T$. \\ \hline
    $J_{\Upsilon,\mathcal{W}}(\gamma,\omega)$ & Joint probability $\mathbb{P}({\Upsilon}>\gamma,\mathcal{W}\le \omega)$. \\ \hline
    \end{tabular}} 
\end{center}
\label{tab:TableOfNotations}
\end{table}

\section{System Model}\label{sec:sys}

This section presents the RIS-assisted network model, association rule, antenna pattern, and channel model.

\subsection{Network Model}\label{subsec:network}

\begin{figure}[t!]
\centering
\vspace{-4mm}
\includegraphics[width=0.9\columnwidth]{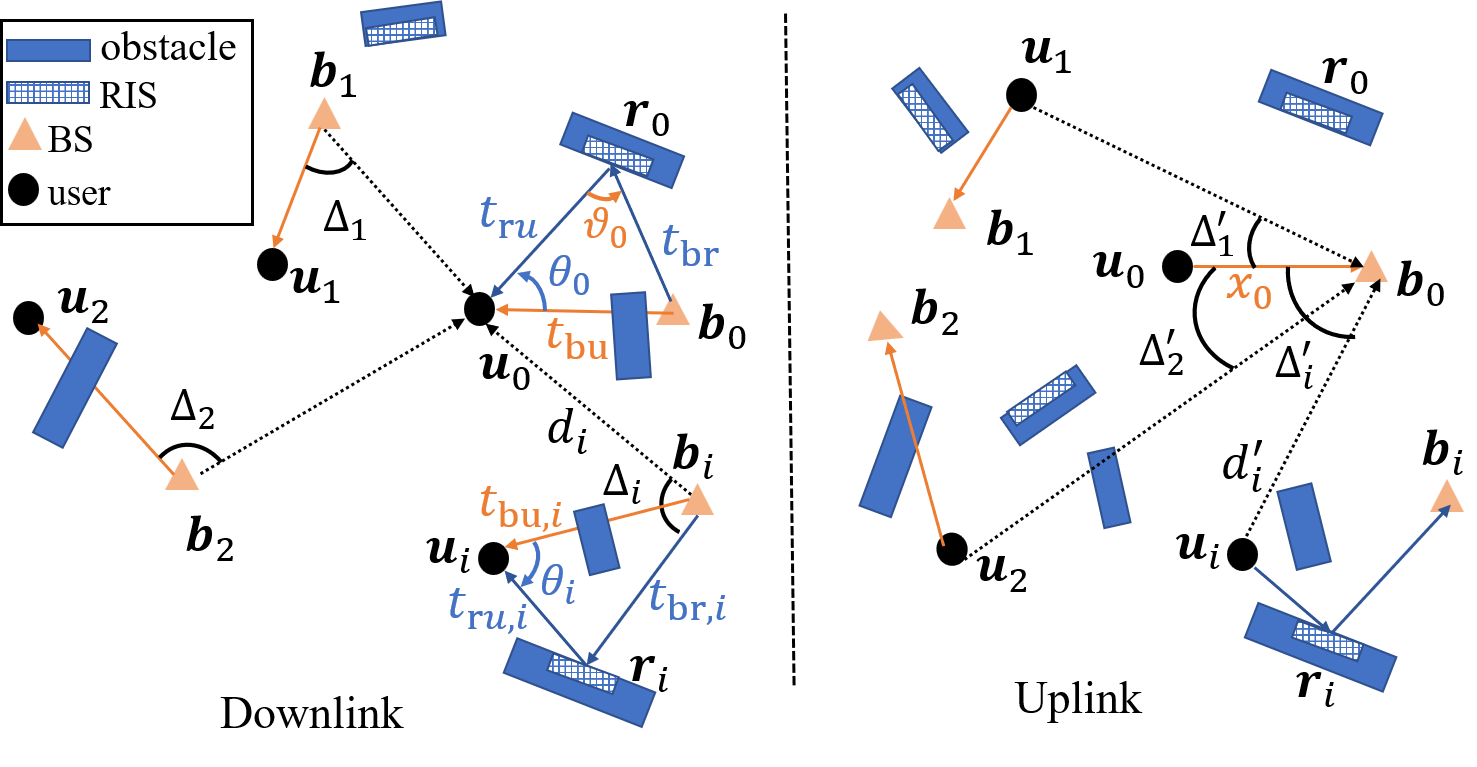}
\caption{{RIS-assisted downlink and uplink communications.}}
\label{fig:sys}
\end{figure}

We consider a \ac{RIS}-assisted cellular network in Fig.~\ref{fig:sys}, where the locations of the $N_{\rm b}$-antenna \acp{BS} and the single-antenna users follow two independent homogeneous \acp{PPP} in $\mathbb{R}^2$: $\Psi_{\rm b}=\{\mathbf{b}_i\}$ with density $\lambda_{\rm b}$ and $\Psi_{\rm u}=\{\mathbf{u}_i\}$ with density $\lambda_{\rm u}$, respectively.
We consider the fully-loaded case of $\lambda_{\rm u} > \lambda_{\rm b}$ and each BS serves one user at each resource block.\footnote{{Our analysis in this paper can be extended to the partially-loaded case of $\lambda_{\rm u} < \lambda_{\rm b}$ based on~\cite{6205422}.}}
Using the line Boolean model, we model the obstacles as line segments with length $\ell$ and orientation $\varphi$. The centers of line segments represent the locations of obstacles, forming a homogeneous \ac{PPP} $\Psi_{\rm o}=\{\mathbf{o}_i\}$ with density $\lambda_{\rm o}$. 
A \ac{RIS} with $N_{\rm r}$ elements is deployed on one side of an obstacle~\cite{kishk2020exploiting}. The obstacles equipped with \acp{RIS} form a subset $\Psi_{\rm r}=\{\mathbf{r}_i\} \subset \Psi_{\rm o}$ with density $\lambda_{\rm r}=\mu\lambda_{\rm o}$, where $0\le \mu\le 1 $ is the fraction of \ac{RIS}-equipped obstacles. 
We define the following three types of serving links between a user and its associated \ac{BS} in the downlink/uplink. 

\begin{definition}[Direct LoS (DL) Link]\label{def:dL}
No obstacle obstructs the straight line between the user and the \ac{BS}.    
\end{definition}
\begin{definition}[Cascaded LoS (CL) Link]\label{def:cL}
\vspace{-1mm}
A cascaded LoS link exists
when a \ac{RIS} satisfies the following two conditions~\cite{kishk2020exploiting}. 
{(i) {\rm \ac{LoS} condition:} there is no obstacle obstructing either the straight line between the \ac{RIS} and the user or between the \ac{RIS} and the \ac{BS}. (ii) {\rm Reflection condition:} both the user and the \ac{BS} are located on the front side of the \ac{RIS}.}
\end{definition}
\begin{definition}[Direct NLoS (DN) Link]\label{def:dN}
\vspace{-1mm}
At least one obstacle obstructs the straight line between the user and the \ac{BS}. Signals can be propagated via multiple paths provided by scatterers. 
\end{definition}
In the following, we use $q$ to indicate the type of the serving link, where $q \in \mathcal{S}\overset{\bigtriangleup }{=} \{\rm DL,CL,DN\}$.

\subsection{Association Rule}\label{subsec:asso}
Generally, a user located at $\mathbf{u}_i$ is associated with its nearest \ac{BS} located at $\mathbf{b}_i$ with {\em \ac{BS}-user distance} $t_{{\rm bu},i}=||\mathbf{b}_i-\mathbf{u}_i||$. 
The \ac{BS} directly transmits (or receives) signals to (or from) the user in the downlink (or uplink) if the direct LoS link exists.
However, in the absence of the direct LoS link, \ac{NLoS} propagation introduces significant attenuation, especially for high-frequency signals, necessitating an alternative association strategy.
We consider that the user establishes a connection with its nearest \ac{RIS} (located at $\mathbf{r}_i$) that can provide a cascaded LoS link between the BS and the user, i.e., this \ac{RIS} satisfies the conditions in Definition~\ref{def:cL}. 
Although this rule may not always be optimal, we adopt it to strike a balance between model practicality and analytical simplicity~\cite{IEEEAccess,10412176}.
The distance between $\mathbf{r}_i$ and $\mathbf{u}_i$ (or $\mathbf{b}_i$) is denoted by {\em \ac{RIS}-user distance} $t_{{\rm ru},i}=||\mathbf{r}_i-\mathbf{u}_i||$ (or {\em \ac{BS}-\ac{RIS} distance} $t_{{\rm br},i}=||\mathbf{b}_i-\mathbf{r}_i||$). 
Moreover, we consider that the link between a \ac{BS} and its connected \ac{RIS} is \ac{LoS}~\cite{distributedRIS,LoSBR}.
If neither a direct LoS nor a cascaded LoS link is available, the direct NLoS link is used for downlink/uplink transmission.

From Slivnyak's theorem~\cite{haenggi2012stochastic}, the statistics observed at a random point of a PPP are the same as those observed at the origin.
Without loss of generality, the following analysis focuses on {\em the typical user} located $\mathbf{u}_0$ and {\em the tagged BS} located at $\mathbf{b}_0$~\cite{andrews2016primer}.
For notation simplicity, we abbreviate $t_{{\rm bu},0}$, $t_{{\rm ru},0}$, and $t_{{\rm br},0}$ as $t_{\rm bu}$, $t_{\rm ru}$, and $t_{\rm br}$, respectively.
As depicted in Fig.~\ref{fig:sys}, by defining the angle from the user-\ac{BS} link to the user-\ac{RIS} link as  {$\theta_0$}, we can express $t_{\rm br}$ as
\begin{align}\label{eq:z0}
    t_{\rm br}=\sqrt{t_{\rm bu}^2+t_{\rm ru}^2-2t_{\rm bu}t_{\rm ru}\cos\theta_0}.
\end{align}


\subsection{Antenna Pattern}\label{subsec:antenna}

A BS equipped with a \ac{ULA} with $N_{\rm b}$ antennas can provide a directional beam via analog beamforming.
Assume that the beam direction of a BS is aligned at the physical AoD $\phi_{\rm p}^*$, which is uniformly distributed, i.e., $\phi_{\rm p}^*\sim \mathcal{U}(-\pi,\pi]$.  We define $\phi_{\rm s}^*= {d}\cos (\phi_{\rm p}^*)/{\lambda_f} $ as the {\em spatial AoD},
where $\lambda_f$ is the wavelength and $d={\lambda_f}/{2}$ is the antenna spacing, i.e., $\phi_{\rm s}^*= \frac{1}{2}\cos (\phi_{\rm p}^*) \in [-\frac{1}{2}, \frac{1}{2}]$.
The beam gain observed at another physical AoD $\phi_{\rm p}\sim \mathcal{U}(-\pi,\pi]$ or spatial AoD $\phi_{\rm s}= \frac{1}{2}\cos (\phi_{\rm p}) $ is~\cite{ActualAntenna}
\begin{align}\label{eq:Gact}
G_{\rm ULA}(\phi_{\rm s}-\phi_{\rm s}^*)=G_{\rm ULA}(\Delta)=\frac{\sin^2(\pi N_{\rm b} \Delta)}{N_{\rm b}\sin^2(\pi\Delta)}.
\end{align}
where $\Delta = \phi_{\rm s}-\phi_{\rm s}^* \in [-1, 1]$ is the spatial AoD deviation off the aligned direction. Note that $G_{\rm ULA}(\cdot)$ is a periodic and even function with $G_{\rm ULA}(\Delta)=G_{\rm ULA}(|\Delta \pm 1|)$.

In the downlink/uplink serving link, a \ac{BS} aligns the beam direction towards its associated user (or \ac{RIS}) in the direct (or cascaded) link to provide the maximum antenna gain  $G_{\rm b}=G_{\rm ULA}(0)=N_{\rm b} $~\cite{RIS_model}.
As for the downlink interfering link from a \ac{BS} at $\mathbf{b}_i$ to the typical user at $\mathbf{u}_0$ ($i\neq 0$), we define $\Delta_i$ as the spatial AoD deviation off the aligned direction $\mathbf{b}_i$-$\mathbf{u}_i$ observed at $\mathbf{b}_i$-$\mathbf{u}_0$. 
The beam gain in the downlink interfering link is $G_{\rm ULA}(\Delta_i)$.
Likewise, the beam gain in the uplink interfering link from the user at $\mathbf{u}_i$ to the tagged BS at $\mathbf{b}_0$ ($i\neq 0$) is $G_{\rm ULA}(\Delta'_i)$, where $\Delta'_i$ is the spatial AoD deviation off the aligned direction $\mathbf{u}_0$-$\mathbf{b}_0$ observed at $\mathbf{u}_i$-$\mathbf{b}_0$, as shown in Fig.~\ref{fig:sys}. 
For analytical tractability in interference characterization, previous works adopted a flat-top antenna pattern to approximate \eqref{eq:Gact}, e.g.,~\cite{IEEEAccess,distributedRIS}. However, the oversimplified flat-top pattern leads to deviations in the network performance analysis~\cite{ActualAntenna}.
To strike a balance between accuracy and tractability, we adopt the following discrete multi-lobe antenna model in~\cite{DiscreteAntenna} for the subsequent interference analysis, i.e.,\footnote{Different from~\cite{ActualAntenna,DiscreteAntenna} where $\Delta$ is assumed to follow a uniform distribution for simplicity, we consider the actual distribution of $\Delta$ based on the uniformly distributed $\phi_{\rm p}^*$ and $\phi_{\rm p}$ in the subsequent analysis.} when $\Delta \in [-\frac{1}{2}, \frac{1}{2}]$,
\begin{align}\label{eq:GB}
G_{\rm B}(\Delta) = \begin{cases} 
G_0,  & \text{ if } |\Delta|\le \frac{1}{N_{\rm b}}, \\
G_n, & \text{ if } \frac{n}{N_{\rm b}}<|\Delta|\le \frac{n+1}{N_{\rm b}},\\
G_{\tilde{N}_{\rm b}+1},  & \text{otherwise},
\end{cases}
\end{align}
where $G_n=\frac{\delta}{2}G_{\rm ULA}(\frac{2n+1}{2N_{\rm b}})$, for $n\in\{1,2,..., \tilde{N}_{\rm b}\}$, $G_0=\frac{\delta}{2}G_{\rm ULA}(0)$, and $G_{\tilde{N}_{\rm b}+1}=0$, where {$\delta$ is a constant compensation factor for roll-off characteristics in \eqref{eq:Gact},} and $\tilde{N}_{\rm b}=\left \lfloor \frac{N_{\rm b}}{2} \right \rfloor -1$.  
Moreover,
for $\Delta\in [-1,-\frac{1}{2})$,
$G_{\rm B}(\Delta)=G_{\rm B}(|\Delta+ 1|)$, and for $\Delta\in (\frac{1}{2},1]$, $G_{\rm B}(\Delta)=G_{\rm B}(|\Delta- 1|)$. The accuracy of the discrete antenna model will be verified in Sec.~\ref{sec:results}.

\subsection{Channel Model}\label{subsec:channel}
By modeling the obstacles as line segments, 
the \ac{LoS} probability of a link with distance $t$ can be expressed as~\cite{Boolean}
\begin{align}\label{eq:PLoS}
\mathcal{P}_{\rm L}(t)=\exp\left(-\beta t \right),
\end{align}
where $\beta=\frac{2\lambda_{\rm o} L_{\rm o}}{\pi}$ and $L_{\rm o} = \mathbb{E}[\ell]$ is the average length of line segments (obstacles).
The corresponding \ac{NLoS} probability is $\mathcal{P}_{\rm N}(t)=1- \mathcal{P}_{\rm L}(t)$. {The path loss is $\zeta t^{-\alpha_{\rm L}}$ or $\zeta t^{-\alpha_{\rm N}}$, where {$\zeta=(\frac{\lambda_f}{4 \pi })^2$} represents the reference path loss 
and $\alpha_v$ ($v\in \{\rm L,N\}$) is the path-loss exponent~\cite{R1-3,R2-3}.} Generally, $\alpha_{\rm N}>\alpha_{\rm L}\ge 2$. We model the small-scale fading as independent Nakagami-m fading with shaping parameter $m_{\rm L}$ ($m_{\rm N}$) for each \ac{LoS} (NLoS) link, {where we assume $m_{\rm L}, m_{\rm N}\in \mathbb{N}$ (integers) for analytical simplicity~\cite{approximate}.}
The \ac{CDF} and \ac{PDF} of the small-scale fading coefficient, denoted by $H_{v}$, $v \in \left \{ \rm L, \rm N \right \}$,
are~\cite{Nakagami}
\begin{subequations}\label{eq:ssfading}
\begin{align}\label{eq:CDF_ssfading}
        F_{H_{v}}(h )&=\frac{\Gamma_l(m_{v},m_{v}h)}{\Gamma(m_{v})}, \\
        \label{eq:PDF_ssfading}
	f_{H_{v}}(h )&=\frac{m_{v}^{m_{v}} h^{m_{v}-1}}{\Gamma(m_{v}) } e^{-m_{v}h },
\end{align}
\end{subequations}
where $\Gamma_l\left ( m,mg \right )=\int_{0}^{mg} x^{m-1}e^{-x}\mathrm{d}x$ is the lower incomplete Gamma function, $\Gamma\left ( m \right )=\int_{0}^{\infty} x^{m-1}e^{-x}\mathrm{d}x$ is the Gamma function, and $\mathbb{E}[H_{v}]=1$. 
Specifically, {in the downlink,} we denote the small-scale fading coefficient in the direct LoS/NLoS link between $\mathbf{b}_0$ and $\mathbf{u}_0$ as $H_{{\rm D}v}$, $v\in\{\rm L,N\}$. $H_{{\rm D}v}$ follows the distribution in \eqref{eq:ssfading} with shaping parameter $m_{{\rm D}v}=m_{v}\in\mathbb{N}$. 
The small-scale fading coefficient in the downlink cascaded LoS link between $\mathbf{b}_0$ and $\mathbf{u}_0$ is denoted by $H_{{\rm CL}}=H_{\rm bu}H_{\rm ru}$, where $H_{\rm bu}$ and $H_{\rm ru}$ are the small-scale fading coefficients in BS-RIS and RIS-user links.
Since the BS-RIS link has strong \ac{LoS}~\cite{distributedRIS,LoSBR}, the shaping parameter of $H_{\rm bu}$ tends to be infinite, i.e., $H_{\rm bu}$ is approximately a constant unit. Therefore, we consider that $H_{{\rm CL}}$ follows the same distribution as $H_{\rm ru}$ with shaping parameter $m_{{\rm CL}}=m_{\rm L}\in\mathbb{N}$ in \eqref{eq:ssfading}~\cite{distributedRIS}.
Similarly, we denote the small-scale fading coefficient in the uplink serving link as $H'_q$, $q\in\mathcal{S}$, with the same distribution as $H_{q}$.

\section{Downlink SINR and \ac{EMFE}}\label{sec:DLmodel}
In this section, we present the modeling of downlink SINR and \ac{EMFE} at the typical user, along with performance metrics to independently characterize the statistics of \ac{SINR} and \ac{EMFE}. 
Moreover, we propose a joint metric to capture their correlation.

\subsection{Downlink SINR}\label{subsec:DLSINR}
This subsection quantifies the downlink SINR by analyzing the serving and interference signal power at the typical user.

\subsubsection{Received Serving Signal Power}
When the serving link is {\em direct LoS/NLoS}, the tagged \ac{BS} at $\mathbf{b}_0$ aligns the beam with the typical user at $\mathbf{u}_0$, providing maximum antenna gain $G_{\rm b}$~\cite{ActualAntenna,RIS_model}. In this case, the received serving signal power at the typical user is
\begin{equation}\label{eq:Pr_dLN}
P_{{\rm D}v}=p_{\rm b} G_{\rm b} \zeta t_{\rm bu}^{-\alpha_{v}} H_{{\rm D}v}=P^{\rm a}_{{\rm D}v}H_{{\rm D}v}, v\in \{\rm L,N\},
\end{equation}
where $p_{\rm b}$ is the \ac{BS} transmit power and $P^{\rm a}_{{\rm D}v}\overset{\bigtriangleup }{=}p_{\rm b} G_{\rm b} \zeta t_{\rm bu}^{-\alpha_{v}}$.
When the serving link is {\em cascaded LoS}, the \ac{BS} at $\mathbf{b}_0$ aligns the beam direction with its associated \ac{RIS} at $\mathbf{r}_0$ to provide maximum antenna gain  $G_{\rm b}$. {By appropriately designing the RIS phases, the \ac{RIS} reflects the incident wave towards the typical user at $\mathbf{u}_0$ with maximum \ac{RIS} gain $G_{\rm r} = N_{\rm r}^2$~\cite{RIS_model,IEEEAccess,distributedRIS}.} 
The corresponding received serving signal power at the typical user is 
\begin{align}\label{eq:Pr_cL}
P_{\rm CL} & =
p_{\rm b} G_{\rm b} G_{\rm r}\zeta (t_{\rm ru}t_{\rm br})^{-\alpha_{\rm L}} H_{{\rm CL}}=P^{\rm a}_{\rm CL}H_{{\rm CL}},
\end{align}
where $P^{\rm a}_{\rm CL}\overset{\bigtriangleup }{=}p_{\rm b} G_{\rm b} G_{\rm r}\zeta (t_{\rm ru}t_{\rm br})^{-\alpha_{\rm L}}$.

\subsubsection{Received Interfering Signal Power}\label{subsec:DLinterference}

The interference experienced by the typical user is dominated by the received interfering signal power from BSs.\footnote{Compared with interference directly introduced by the interfering \acp{BS}, the signal power reflected by the interfering \acp{RIS} with phase shifts optimized for their own serving user (instead of the typical user) is relatively weak due to the increased path loss through cascaded links~\cite{tutorialRIS,IEEEAccess}.}
A BS located at $\mathbf{b}_i$ may interfere with the typical user at $\mathbf{u}_0$ via a LoS (or NLoS) link with probability $\mathcal{P}_{\rm L}(d_i)$ (or $\mathcal{P}_{\rm N}(d_i)$), where $d_i=||\mathbf{b}_i-\mathbf{u}_0||$. Therefore, we have $\Psi_{\rm b}=\Psi_{\rm b,L}\cup \Psi_{\rm b,N}$, where the link between the \ac{BS} at $\mathbf{b}_i \in \Psi_{\rm b,L}$ (or $\Psi_{\rm b,N}$) and the typical user is \ac{LoS} (or \ac{NLoS}).
The density of $\Psi_{{\rm b},v}$ is $\lambda_{\rm b}\mathcal{P}_{v}(d_i)$, where $v\in\{\rm L,N\}$.
The interference from BSs in $\Psi_{{\rm b},v}$, ${v\in \{\rm L,N\}}$, is 
\begin{align}\label{eq:Ibv}
	I_{{\rm b},v}=\sum_{i,\mathbf{b}_i\in\Psi_{{\rm b},v}\setminus \{\mathbf{b}_0\}}
	{p_{\rm b} G_{\rm B}(\Delta_{i})\zeta d_i ^{-\alpha_{v}} H_{v,i}}, 
\end{align}
where 
$G_{\rm B}(\Delta_i)$ is antenna gain of the \ac{BS} at $\mathbf{b}_i$, $d_i=||\mathbf{b}_i-\mathbf{u}_0||$, $H_{v,i}$ is the small-scale fading coefficient of the interfering link, and $H_{v,i}$ follows the distribution in \eqref{eq:ssfading}.
Therefore, the aggregate interference from BSs is $I_{\rm B}=I_{\rm b,L}+I_{\rm b,N}$.

\subsubsection{Downlink SINR and Coverage Probability}

The probability that the downlink \ac{SINR} ($\Upsilon$) is above a predefined threshold ($\gamma$)
is defined as the {\em downlink coverage probability}, denoted by $\bar F_{\Upsilon}(\gamma)$.
Namely, $\bar F_{\Upsilon}(\gamma)$ is the \ac{CCDF} of $\Upsilon$, i.e.,
\begin{align}\label{eq:cov_def}
\bar F_{\Upsilon}(\gamma) & = \mathbb{P}\left ({\Upsilon}>\gamma \right ).
\end{align}
Considering that the serving link may be direct LoS, cascaded LoS, or direct NLoS, the corresponding downlink \ac{SINR}, denoted by $\Upsilon_q$, is
\begin{align}\label{eq:SINRq}
{\Upsilon}_{q}=\frac{P_{q}}{\sigma^2+I_{\rm B}}, q\in \mathcal{S},
\end{align}
where $\sigma^2$ is the noise power in the downlink.
Then, based on the total probability law, we can express \eqref{eq:cov_def} as
\begin{align}\label{eq:CCDF}
\bar F_{\Upsilon}(\gamma) &= \sum_{q\in\mathcal{S}} \mathbb{P}\left  ( {\Upsilon_q}>\gamma \right ) \mathcal{A}_{q},
\end{align}
where $\mathcal{A}_{q}$ is the association probability, e.g., $\mathcal{A}_{\rm DL}$ is the probability that the typical user is associated with the tagged BS via the direct LoS link, which will be analyzed in Sec.~\ref{subsec:assoProb}.

\subsection{Downlink \ac{EMFE}}\label{subsec:DLEMF}

In the \ac{RIS}-assisted downlink communication, 
\ac{EMFE} is introduced by \ac{BS} transmission and \ac{RIS} reflection, which can be quantified by the received power density at the typical user~\cite{Joint}.

\subsubsection{Received Power Density}
Based on the received serving signal power $P_{q}$ given in \eqref{eq:Pr_dLN} and \eqref{eq:Pr_cL}, the received serving signal power density at the typical user is
\begin{align}\label{eq:Wb0}
\mathcal{W}_{1,q} & = \frac{P_{q}}{\mathcal{E}}=\frac{P^{\rm a}_{q} H_q}{\mathcal{E}},
\end{align}
where $q\in \mathcal{S}$ is the serving link type, 
$\mathcal{E} = \frac{\lambda_f^2 G_{\rm u}}{4\pi}$ is the antenna effective area of the typical user, and $G_{\rm u}\!=\!1$ for the single-antenna user.
Similarly, from \eqref{eq:Ibv}, the received interfering signal power density at the typical user is
\begin{align}\label{eq:W2}
\mathcal{W}_{2}&=\frac{I_{\rm B}}{\mathcal{E}}=\frac{I_{\rm b,L}+I_{\rm b,N}}{\mathcal{E}}=\mathcal{W}_{\rm b,L} +\mathcal{W}_{\rm b,N}, 
\end{align}
where $\mathcal{W}_{{\rm b},v}\overset{\bigtriangleup }{=} \frac{I_{{\rm b},v}}{\mathcal{E}}$.
Therefore, considering the type of the serving link $q\in \mathcal{S}$, the aggregate downlink \ac{EMFE}, denoted by $\mathcal{W}_q~\left [ \rm W/m^2 \right ]$, is
\begin{align}\label{eq:Wq}
\mathcal{W}_{q}&=\mathcal{W}_{1,q}+\mathcal{W}_{2}=\frac{P_{q}+I_{\rm B}}{\mathcal{E}}.
\end{align}

\subsubsection{Downlink \ac{EMFE} Constraint and Compliance Probability}
The probability that the downlink \ac{EMFE} ($\mathcal{W}$) is below a constraint level ($\omega$), i.e., the \ac{CDF} of \ac{EMFE}, is defined as the {\em downlink compliance probability}, denoted by $F_\mathcal{W}(\omega)$.
The lower the \ac{EMFE}, the higher the compliance probability.
Similar to \eqref{eq:CCDF}, the downlink compliance probability can be written as 
\begin{align}\label{eq:CDF_W}
F_\mathcal{W}(\omega) & = \mathbb{P}(\mathcal{W}\le \omega) = \sum_{q\in\mathcal{S}} \mathbb{P}({\mathcal{W}_{q}}\le \omega)\mathcal{A}_{q}.
\end{align}
Regulatory authorities have imposed a constraint on \ac{EMFE}, requiring it to remain below the maximum allowed value $\mathcal{W}_{\max}$ with high probability $\rho$ \cite{ITU5G:19}, i.e., $F_{\mathcal{W}}(\mathcal{W}_{\max})\geq\rho$.
We can express the {\em \ac{EMFE} constraint} as
\begin{equation}\label{eq:inverseF}
    F^{-1}_{\mathcal{W}}(\rho) \leq \mathcal{W}_{\max},
\end{equation}
where $F^{-1}_{\mathcal{W}}(\rho)$ is the inverse function of $F_{\mathcal{W}}(\omega)$ in \eqref{eq:CDF_W}.
From~\cite{FCCEvaluatingCom:97},  $\mathcal{W}_{\max} = 10~\rm  W/m^2$ and $\rho=0.95$. 
In particular, we define $F^{-1}_{\mathcal{W}}(0.95)$ as {\em 95-th percentile of \ac{EMFE}}, which means the probability that the \ac{EMFE} exceeds $F^{-1}_{\mathcal{W}}(0.95)$ is below 5\%.

\subsection{Joint Metric of Downlink SINR and \ac{EMFE}}
From the above discussion, we observe that the statistical characteristics of downlink SINR ($\Upsilon$) and \ac{EMFE} ($\mathcal{W}$) are correlated via the distance, the small-scale fading, and the type of the serving link. Therefore, instead of independently analyzing the downlink SINR or \ac{EMFE} distribution by the downlink coverage or compliance probability, we define a joint metric as follows. 
\begin{definition}[Joint Coverage and Compliance Probability]\label{def:joint}
The probability that both (i) \ac{SINR} is above a predefined threshold, and (ii) \ac{EMFE} is below a constraint level, is defined as the joint coverage and compliance probability, abbreviated as joint probability.
A high joint probability is achieved when SINR is high and \ac{EMFE} is low.
\end{definition}
The downlink joint probability, denoted by $J_{\Upsilon,\mathcal{W}}(\gamma,\omega)$, is
\begin{align}\label{eq:joint}
\begin{split}
J_{\Upsilon,\mathcal{W}}(\gamma,\omega) &=\mathbb{P}({\Upsilon}>\gamma,\mathcal{W}\le \omega) \\& =\sum_{q\in\mathcal{S}}
\mathbb{P}(\Upsilon_q>\gamma,\mathcal{W}_q\le \omega)\mathcal{A}_q.     
\end{split}
\end{align} 
From Definition~\ref{def:joint}, the joint probability allows for a quantified assessment of both SINR and \ac{EMFE}. 
This is particularly useful in cases where deployment strategies are designed to improve SINR while potentially exacerbating \ac{EMFE}.
Using the joint probability, alternative strategies can be explored to balance SINR and \ac{EMFE}.

\section{Downlink Performance Analysis}\label{sec:DLanaly}

This section derives the expressions for the downlink performance metrics.  
We first provide the association probabilities and the distance distributions of different serving links.    
Then, we derive the downlink coverage and compliance probabilities independently, i.e., the marginal distributions of $\Upsilon$ and $\mathcal{W}$. {We also derive the first moment of $\mathcal{W}$.} Based on the \ac{EMFE} distribution, we investigate the \ac{CD}.
Finally, we derive the joint distribution of $\Upsilon$ and $\mathcal{W}$.

\subsection{Association Probability}\label{subsec:assoProb}

\subsubsection{Association with a Direct \ac{LoS} Link}
From Sec.~\ref{subsec:asso}, the probability that the typical user located at $\mathbf{u}_0$ is associated with the \ac{BS} located at $\mathbf{b}_0$ via the direct \ac{LoS} link is the \ac{LoS} probability of the direct link, i.e., $\mathcal{A}_{\rm DL}=\mathcal{P}_{\rm L}(t_{\rm bu})$.

\subsubsection{Association with a Cascaded \ac{LoS} Link} 
Recall that there are \textcolor{black}{two independent conditions} of the cascade LoS link in Definition~\ref{def:cL}. For a RIS located at $\mathbf{r}$, a cascaded LoS link between the typical user at $\mathbf{u}_0$  and the tagged BS at $\mathbf{b}_0$ exists if the RIS satisfies (i) \ac{LoS} condition of the \ac{BS}-\ac{RIS} and \ac{RIS}-user links and (ii) reflection condition.
\textcolor{black}{Considering that the \ac{BS}-\ac{RIS} link is strong LoS~\cite{distributedRIS,LoSBR}, the probability that the RIS satisfies condition (i) is the LoS probability of the user-\ac{RIS} link, i.e., $\mathcal{P}_{\rm L}(t)$, where $t=||\mathbf{r}-\mathbf{u}_0||$.}  The probability of the RIS satisfying condition (ii) is~\cite[Appendix A]{kishk2020exploiting}  
\begin{align}\label{eq:refProb}
\mathcal{P}_{\rm R}(t_{\rm bu},t,\theta)\!=\!\!\frac{1}{2}\!-\!\frac{1}{2\pi}\arccos\!\left(\!\!\frac{t -t_{\rm bu}\cos\theta}{\sqrt{t_{\rm bu}^2\!+\!t^2\!-\!2t_{\rm bu}t\cos\theta}}\!\!\right)\!,
\end{align}
where $\theta$ is the angle from the user-\ac{BS} link to the user-\ac{RIS} link. 
Then, the probability that the \ac{RIS} at $\mathbf{r}$ can create a cascaded LoS link between $\mathbf{u}_0$ and $\mathbf{b}_0$ is
\begin{align}\label{eq:existence}
&a(t_{\rm bu},t,\theta)=\mathcal{P}_{\rm L}(t_{\rm ru})\mathcal{P}_{\rm R}(t_{\rm bu},t,\theta).
\end{align}
To establish a cascaded LoS link between $\mathbf{u}_0$ and $\mathbf{b}_0$, it is necessary that at least one \ac{RIS} in $\Psi_{\rm r}$ satisfies the above two conditions. Therefore, the probability of the existence of a cascaded LoS link between $\mathbf{u}_0$ and $\mathbf{b}_0$ is~\cite[Lemma 2]{kishk2020exploiting}
\begin{align}\label{eq:PcL} 
\mathcal{P}_{\rm CL}(t_{\rm bu})=1-\exp\left(-\lambda_{\rm r} \int_0^\infty  \bar a (t_{\rm bu},t) t\,{\rm d}t \right), 
\end{align}
where $\bar a (t_{\rm bu},t)= \int_{-\pi}^{\pi}a(t_{\rm bu},t ,\theta) \mathrm {d}\theta $. As discussed in Sec.~\ref{subsec:asso}, the serving link is cascaded LoS if the direct LoS link is unavailable and if the cascaded LoS link exists. Hence, 
we have $\mathcal{A}_{\rm CL}=\mathcal{P}_{\rm N}(t_{\rm bu})\mathcal{P}_{\rm CL}(t_{\rm bu})$.

\subsubsection{Association with a Direct \ac{NLoS} Link}
The probability of associating with a direct NLoS link is $\mathcal{A}_{\rm DN}=1-\mathcal{A}_{\rm DL}-\mathcal{A}_{\rm CL}=\mathcal{P}_{\rm N}(t_{\rm bu})(1-\mathcal{P}_{\rm CL}(t_{\rm bu}))$.

\subsection{Beam Gain Distribution}\label{subsec:angleDist}
With the discrete antenna model $G_{\rm B}(\Delta)$ in \eqref{eq:GB}, we discretize the antenna gain as a constant $G_n$, $n\in\{0,1,..., \tilde{N}_{\rm b}\}$, at different intervals of $\Delta$. For the subsequent interference analysis, we provide the probability of the antenna gain being $G_n$ based on the distribution of $\Delta$ in the following.

We denote the \ac{PDF} of $\Delta$ as $f_{\Delta}(\cdot)$. As defined in Sec.~\ref{subsec:antenna}, $\Delta=\phi_{\rm s}-\phi_{\rm s}^*= \frac{1}{2}\cos (\phi_{\rm p}) - \frac{1}{2}\cos (\phi_{\rm p}^*)$.  
Based on the uniform distribution of the physical AoDs $\phi_{\rm p}$ and $\phi_{\rm p}^*$ in $(-\pi,\pi]$, the \ac{pdf} of $\Delta$ is 
\begin{align}\label{eq:pdf_Delta}
f_{\Delta}(y) =
\begin{cases}
\int_{-\frac{1}{2}}^{\frac{1}{2}} f_{\Phi_{\rm s}}(x)f_{\Phi_{\rm s}}(y+x) \mathrm{d} x, & \!\!\text{for } y \in [-1,1],\\
 0, & \!\!\text{otherwise},
\end{cases}
\end{align}
where $f_{\Phi_{\rm s}}(\cdot)$ is the \ac{pdf} of $\phi_{\rm s}$ and $\phi_{\rm s}^*$. Based on the uniform distribution of $\phi_{\rm p}$, we have $f_{\Phi_{\rm s}}(x)=\frac{2}{\pi\sqrt{1-4x^2}}$ for $x \in [-\frac{1}{2},\frac{1}{2}]$.
\begin{lemma}\label{lemma:prob_Gain}
Under the discrete antenna model $G_{\rm B}(\Delta)$ in \eqref{eq:GB}, the probability that the beam gain is $G_n$ is given by
\begin{align}\label{eq:pn}
p_n & = 2 \int_{\frac{n}{N_{\rm b}}}^{\frac{n+1}{N_{\rm b}}} f_{\Delta}(y)\mathrm{d} y + 2 \int_{1-\frac{n+1}{N_{\rm b}}}^{1-\frac{n}{N_{\rm b}}}
f_{\Delta}(y)\mathrm{d} y,
\end{align}
where $n\in\{0,1,..., \tilde{N}_{\rm b}\}$ and $\tilde{N}_{\rm b}=\left \lfloor \frac{N_{\rm b}}{2} \right \rfloor -1$.
The probability that the beam gain is $G_{\tilde{N}_b+1}=0$ is  $p_{\tilde{N}_b+1}=1-\sum_{n=0}^{\tilde{N}_b} p_n$.
\begin{IEEEproof}
From \eqref{eq:GB}, when $\Delta\in (\frac{n}{N_{\rm b}},\frac{n+1}{N_{\rm b}}] \cup [1-\frac{n+1}{N_{\rm b}}, 1-\frac{n}{N_{\rm b}}) \cup (\frac{n}{N_{\rm b}}-1, \frac{n+1}{N_{\rm b}}-1] \cup [-\frac{n+1}{N_{\rm b}},-\frac{n}{N_{\rm b}})$, we have $G_{\rm B}(\Delta)=G_n$. Based on $f_{\Delta}(\cdot)$ in \eqref{eq:pdf_Delta}, we obtain \eqref{eq:pn}. 
\end{IEEEproof}
\end{lemma}
Note that $p_n$ depends on  the antenna number $N_{\rm b}$. Given $N_{\rm b}$, $p_n$ can be numerically calculated offline, which will greatly simplify the subsequent interference analysis.

\subsection{Serving Link Distance Distribution}\label{subsec:dist}

From Sec.~\ref{subsec:asso}, the typical user is served by its nearest BS.
Based on the null probability of a PPP,
the \ac{PDF} of the \ac{BS}-user distance $t_{\rm bu}$ in the direct link is~\cite{andrews2016primer}
\begin{align}\label{eq:fX}
f_{T_{\rm bu}}(t_{\rm bu}) & = 2\pi\lambda_{\rm b} t_{\rm bu}\exp(-\pi\lambda_{\rm b}t_{\rm bu}^2).
\end{align}
We provide the distribution of \ac{RIS}-user distance $t_{\rm ru}$ in the following lemma.
\begin{lemma}\label{lemma:cas_distance}
Conditioned on the \ac{BS}-user distance $t_{\rm bu}$,
the \ac{PDF} of the \ac{RIS}-user distance $t_{\rm ru}$  in the cascaded link is
\begin{align}
&f_{T_{\rm ru}|t_{\rm bu}}(t_{\rm ru})  = \frac{\lambda_{\rm r} t_{\rm ru} \bar a (t_{\rm bu},t_{\rm ru})
\exp(-\lambda_{\rm r} \int_{0}^{t_{\rm ru}} \bar a (t_{\rm bu},t) t \mathrm{d}t)}{\mathcal{P}_{\rm CL}(t_{\rm bu})}.
\end{align} 
\begin{IEEEproof}
Based on the reflection probability in \eqref{eq:existence} and the void probability of a \ac{PPP}, we can finish the derivation~\cite{wcnc}.
\end{IEEEproof}
\end{lemma}
Given $t_{\rm ru}$ and $t_{\rm bu}$, the distribution of \ac{BS}-\ac{RIS} distance $t_{\rm br}$ is dependent on $\theta_0\sim\mathcal{U}(-\pi,\pi]$, as shown in \eqref{eq:z0}.

\subsection{Downlink SINR Analysis}\label{subsec:cov}

Let $\mathcal{T}_q$ be the set of random variables relevant to the distances of different types of the serving link, i.e., $\mathcal{T}_{\rm DL}=\mathcal{T}_{\rm DN}=\{t_{\rm bu}\}$, $\mathcal{T}_{\rm CL}=\{t_{\rm bu},t_{\rm ru},t_{\rm br}\}$. 
Then, we can define the conditional coverage probability as the \ac{CCDF} of ${\Upsilon}_{q}$ conditioned on $\mathcal{T}_q$, denoted by $\bar F_{{\Upsilon}_{q}|\mathcal{T}_q}(\gamma)$.  {For Nakagami-m parameter $m_q\in \mathbb{N}$, $q\in\mathcal{S}$, we can express $\bar F_{{\Upsilon}_{q}|\mathcal{T}_q}(\gamma)$ as~\cite[Theorem 1]{approximate}}
\begin{align}\label{eq:ccdf_q}
&\bar F_{{\Upsilon}_{q}|\mathcal{T}_q}(\gamma)=\mathbb{P}({\Upsilon}_{q}>\gamma|\mathcal{T}_q) 
\nonumber\\&\overset{(a)}{=}\sum_{k=0}^{m_{q}-1} \frac{(-s_{q})^k}{k!} \frac{\partial^k \exp(- s_{q}\sigma^2)\mathcal{L}_{I_{\rm B}|t_{\rm bu}}( s_{q} )}{\partial s_{q}^k}   
\nonumber\\&\overset{(b)}{\approx}\sum_{k=1}^{m_{q}}\!\binom{m_{q}}{k} (-1)^{k+1} \!\exp(-k \beta_{q} s_{q}\sigma^2) \mathcal{L}_{I_{\rm B}|t_{\rm bu}}( k \beta_{q} s_{q} ) ,
\end{align} 
where
$s_{q}= \frac{m_{q}\gamma}{P^{\rm a}_{q}}$, 
$P^{\rm a}_q$ is given in \eqref{eq:Pr_dLN} and \eqref{eq:Pr_cL},
$\beta_{q}  = \left ( m_{q}!\right ) ^{\frac{-1}{m_{q}} }$, $\mathcal{L}_{I_{\rm B}|t_{\rm bu}}(s)=\mathbb{E}\left[\exp\left(-sI_{\rm B}\right)\right]$ is derived in the following lemma, the calculation in (a) would be quite complex due to the high order of derivations of the Laplace transform, and (b) provides an approximate expression to simplify the computation by using the upper bound of the CDF of the Gamma distribution~\cite[Appendix F]{approximate}.

\begin{lemma}\label{lemma:L_SINR}
The Laplace transform of $I_{\rm B}$ conditioned on $t_{\rm bu}$ is 
\begin{align}\label{eq:LIB0}
\mathcal{L}_{I_{{\rm B}}|t_{\rm bu}} (s) &=\!\!\!\prod_{v\in \{\rm L,N\}}\!\!\exp\bigg(\!\!-2\pi\lambda_{\rm b}\times
\nonumber\\&\int_{t_{\rm bu}}^{\infty}\!\!\!
[1-\kappa_{v} ( {s p_{\rm b} \zeta }{d_i ^{-\alpha_{v}}})] d_i \mathcal{P}_{v}(d_i)
\mathrm{d} d_i
\bigg),
\end{align}
where $\kappa_{v} (x)=\sum_{n=0}^{\tilde{N}_{\rm b}+1}
p_n\left(1+ \frac{x G_n}{m_{v} }\right)^{-m_{v}}$,
$G_n$ is given in \eqref{eq:GB}, $p_n$ is given in \eqref{eq:pn}, and $\mathcal{P}_{v}(\cdot)$ is the LoS (or NLoS) probability for $v={\rm L}$ (or $v={\rm N}$).
\begin{IEEEproof}
See Appendix~\ref{app:lemma:L_SINR}.
\end{IEEEproof}
\end{lemma} 
With the conditional coverage probability $\bar F_{{\Upsilon}_{q}|\mathcal{T}_q}(\gamma)$, we derive the downlink coverage probability as follows.
\begin{theorem}[Downlink Coverage Probability]\label{theorem:cov}
The \ac{CCDF} of the downlink \ac{SINR} ($\Upsilon$) in \eqref{eq:CCDF} is 
 \begin{align} \label{eq:cov}
&\bar F_{\Upsilon}(\gamma)  =\mathbb{E}_{t_{\rm bu},t_{\rm ru},t_{\rm br}} \bigg[ \sum_{q\in\mathcal{S}}\mathcal{A}_{q} \bar F_{{\Upsilon}_{q}|\mathcal{T}_{q}} (\gamma) \bigg]
\nonumber\\&\overset{(a)}{=}\mathbb{E}_{t_{\rm bu},t_{\rm ru},\theta_0} \bigg[ \sum_{q\in\mathcal{S}}\mathcal{A}_{q} \bar F_{{\Upsilon}_{q}|\mathcal{T}_{q}} (\gamma) \bigg]
=\int_{0}^{\infty}\bigg[\mathcal{A}_{\rm DL}\bar F_{{\Upsilon}_{\rm DL}|t_{\rm bu}} (\gamma)
\nonumber\\&+\int_{0}^{\infty}\int_{-\pi}^{\pi}\mathcal{A}_{\rm CL} \bar F_{{\Upsilon}_{\rm CL}|t_{\rm bu},t_{\rm ru},\theta_0} (\gamma)
{\frac{1}{2\pi} f_{T_{\rm ru}|t_{\rm bu}}(t_{\rm ru})  \mathrm{d}\theta_0 \mathrm{d}t_{\rm ru}}
\nonumber\\&+\mathcal{A}_{\rm DN}\bar F_{{\Upsilon}_{\rm DN}|t_{\rm bu}} (\gamma)
\bigg] f_{T_{\rm bu}}(t_{\rm bu}) \mathrm{d}t_{\rm bu},
\end{align}
where (a) is from \eqref{eq:z0}, and $\bar F_{{\Upsilon}_{\rm CL}|t_{\rm bu},t_{\rm ru},\theta_0}$ is obtained by substituting \eqref{eq:z0} into \eqref{eq:ccdf_q}.
\end{theorem}
To simplify the expressions in the subsequent analysis, we do not expand the expectation operation over $t_{\rm bu}$, $t_{\rm ru}$, and $t_{\rm br}$, i.e., $\mathbb{E}_{t_{\rm bu},t_{\rm ru},t_{\rm br}}[\cdot]$ can be conducted in the same way as \eqref{eq:cov}.

\subsection{Downlink \ac{EMFE} Analysis}\label{subsec:EMF}

The \ac{CDF} of $\mathcal{W}_{q}$ conditioned on $\mathcal{T}_q$, i.e., conditional compliance probability, denoted by $F_{\mathcal{W}_{q}|\mathcal{T}_q}(\omega)$, $q\in\mathcal{S}$, is given by~\cite[Appendix C]{Joint}
\begin{align} \label{eq:F_Wq}
&F_{\mathcal{W}_{q}|\mathcal{T}_q}(\omega)=\mathbb{P}(\mathcal{W}_{q}\leq \omega|\mathcal{T}_q) 
\nonumber\\&\overset{(a)}{=}
\frac{1}{2} - \int_{0}^{\infty} \frac{1}{\pi x} {\rm Im} \big[ e^{-{j} x \omega}\mathcal{L}_{\mathcal{W}_{q}|\mathcal{T}_q}(-{j} x)\big ] {\rm d} x.
\end{align}
where (a) is from Gil-Pelaez's inversion theorem~\cite{inversion}, ${\rm Im}[s]$ is the imaginary part of $s$, $j$ is the imaginary unit with $j^2=-1$, and 
$\mathcal{L}_{\mathcal{W}_{q}|\mathcal{T}_q}(s)=\mathbb{E}\left [  \exp(-s \mathcal{W}_{q}) \right ]$ is derived below.
\begin{lemma}\label{lemma:emf_dirL}
The Laplace transform of $\mathcal{W}_{q}$ conditioned on $\mathcal{T}_q$ is 
\begin{align} \label{eq:L_WcL}
\mathcal{L} _{\mathcal{W}_{q}|\mathcal{T}_q}(s)
& =\left(1+ \frac{s P^{\rm a}_{q} }{m_{q} \mathcal{E}}\right)^{-m_{q}}\
\!\!\!\mathcal{L} _{\mathcal{W}_{2}|t_{\rm bu}}(s),
\end{align}
where  
$\mathcal{L}_{\mathcal{W}_{2}|t_{\rm bu}}(s)  = \mathcal{L}_{\mathcal{W}_{\rm b,L}|t_{\rm bu}}(s) \mathcal{L}_{\mathcal{W}_{\rm b,N}|t_{\rm bu}}(s) $,
\begin{align}\nonumber
&\mathcal{L}_{\mathcal{W}_{{\rm b},v}|t_{\rm bu}}(s)
\!=\!\exp\!\left(\!  
-\!2\pi\lambda_{\rm b}\!\!\int_{t_{\rm bu}}^{\infty}\!\! 
\left[1\!-\!\kappa_{v} \left(\!\frac{s p_{\rm b}}{4 \pi d_i ^{\alpha_{v}}}\!\right)\right] d_i \mathcal{P}_{v}(d_i)
\mathrm{d} d_i\!\!
\right),
\end{align} 
$v\in\{L,N\}$, and $\kappa_{v}(\cdot)$ is given in Lemma~\ref{lemma:L_SINR}.
\begin{IEEEproof}
From \eqref{eq:Wq}, we have 
\begin{align}
\mathcal{L} _{\mathcal{W}_{q}|\mathcal{T}_q}(s)
&=\mathbb{E}\left [  \exp(-s \mathcal{W}_{q}) \right ] 
=\mathbb{E}\left [ \exp\left (-s (\mathcal{W}_{1,q}+\mathcal{W}_{2})\right )  \right ] 
\nonumber\\&=\mathbb{E}\left [ \exp\left (-s \mathcal{W}_{1,q}\right )  \right ]\times
\mathbb{E}\left [ \exp\left (-s \mathcal{W}_{2}\right )  \right ]     
\nonumber\\&\overset{(a)}{=}\left(1+ \frac{s P^{\rm a}_{q} }{m_{q} \mathcal{E}}\right)^{-m_{q}}\
\!\!\!\mathcal{L} _{\mathcal{W}_{2}|t_{\rm bu}}(s) ,
\end{align}
where (a) is from \eqref{eq:ssfading}  and $\mathcal{L} _{\mathcal{W}_{2}|t_{\rm bu}}(s) \triangleq \mathbb{E}\left [ \exp\left (-s \mathcal{W}_{2}\right )  \right ] $.
We can derive $\mathcal{L} _{\mathcal{W}_{2}|t_{\rm bu}}(s)$ using the same methods in Appendix~\ref{app:lemma:L_SINR}, which is omitted here.
\end{IEEEproof}
\end{lemma}

By averaging \eqref{eq:F_Wq} over distances $t_{\rm bu}$, $t_{\rm ru}$, or $t_{\rm br}$ with the distributions derived in Sec.~\ref{subsec:dist},
{we obtain the \ac{CDF} of the \ac{EMFE} $\mathcal{W}$ in \eqref{eq:CDF_W} and $\mathbb{E}[\mathcal{W}]$ in the following theorem.}

\begin{theorem}[Downlink \ac{EMFE}]\label{theorem:emf} The \ac{CDF} of overall \ac{EMFE} $\mathcal{W}$ in \eqref{eq:CDF_W} is
\begin{align}\label{eq:emf}
&F_{\mathcal{W}}(\omega)  = \mathbb{E}_{t_{\rm bu},t_{\rm ru},t_{\rm br}} \bigg[ \sum_{q\in\mathcal{S}}\mathcal{A}_{q} F_{\mathcal{W}_{q}|\mathcal{T}_{q}} (\omega) \bigg],
\end{align}
where $\mathcal{A}_q$ is given in Sec.~\ref{subsec:assoProb}, $F_{\mathcal{W}_{q}|\mathcal{T}_{q}} (\omega)$ is given in \eqref{eq:F_Wq}, and $\mathbb{E}_{t_{\rm bu},t_{\rm ru},t_{\rm br}}[\cdot]$ can be conducted in the same way as \eqref{eq:cov}. 
Moreover, the first moment of $\mathcal{W}$ is 
$\mathbb{E}[\mathcal{W}]  
= \mathbb{E}[\mathcal{W}_1]+ \mathbb{E}[\mathcal{W}_2]$, where $\mathbb{E}[\mathcal{W}_1]$ and $\mathbb{E}[\mathcal{W}_2]$ are related to the received serving and interfering signal power density, respectively, i.e.,   
\begin{subequations}\label{eq:E_W}
\begin{align}\label{eq:E_W1}
       \mathbb{E}[\mathcal{W}_1]&=\mathbb{E}_{t_{\rm bu},t_{\rm ru},t_{\rm br}}\bigg[ \sum_{q\in\mathcal{S}} \frac{P^{\rm a}_{q}}{\mathcal{E}} \mathcal{A}_{q}\bigg], \\
        \label{eq:E_W2}
\mathbb{E}[\mathcal{W}_2] &=\frac{\lambda_{\rm b}  p_{\rm b} \bar g_{\rm B} }{2} \bigg(\mathbb{E}_{t_{\rm bu}}\bigg[ 
   t_{\rm bu}^{2-\alpha_{\rm L}} {\rm E}_{\alpha_{\rm L}-1}(\beta t_{\rm bu})
\bigg]+
\nonumber\\& \quad \mathbb{E}_{t_{\rm bu}}\bigg[ 
  \frac{t_{\rm bu}^{2-\alpha_{\rm N}}}{\alpha_{\rm N}-2} -  t_{\rm bu}^{2-\alpha_{\rm N}} {\rm E}_{\alpha_{\rm N}-1}(\beta t_{\rm bu})      
\bigg]  \bigg),
\end{align}
\end{subequations}
where $P^{\rm a}_{q}$ is given in \eqref{eq:Pr_dLN} and \eqref{eq:Pr_cL},  $\bar g_{\rm B} = \mathbb{E}_{\Delta} [G_{\rm B}(\Delta)] =\sum_{n=0}^{\tilde{N}_{\rm b}} G_np_n$ is from Lemma~\ref{lemma:prob_Gain}, $\alpha_{\rm N}>\alpha_{\rm L}\ge 2$,
$\beta=\frac{2\lambda_{\rm o} L_{\rm o}}{\pi}$ is defined in \eqref{eq:PLoS}, ${\rm E}_n(x) = \int_{1}^{\infty}
 \frac{\exp\left(-x t \right) }{t ^{n} } \mathrm{d} t $ can be evaluated by a built-in function ${\rm expint} (n,x)$ in Matlab. 
 \begin{IEEEproof}
See Appendix~\ref{app:theorem:emf}.
\end{IEEEproof}
\end{theorem}

The first-moment analysis of downlink EMFE reveals that a lower BS density ($\lambda_{\rm b}$) decreases the average EMFE from interfering links, i.e., $\mathbb{E}[\mathcal{W}_2]$, due to the reducing number of radiating sources.
Moreover, by noting that $\mathbb{E}[\mathcal{W}_2]$ is primarily influenced by its first term in \eqref{eq:E_W2} (i.e., received interfering power density via LoS links) and ${\rm E}_n(x)$ is a decreasing function, we find that a higher obstacle density ($\lambda_{\rm o}$), i.e., larger $\beta$, results in a lower value of $\mathbb{E}[\mathcal{W}_2]$. These insights underscore the less significant impact of received interfering power density in \ac{EMFE}, especially in environments with dense obstacles or sparse BSs.

\subsection{Compliance Distance} \label{subsec:Xcom}

The \ac{CD} around a BS, denoted by {$\tau_{\rm bu}$},
represents the minimum allowable separation between a \ac{BS} and a user, ensuring compliance with the \ac{EMFE} constraint in \eqref{eq:inverseF}. 
Specifically, {if the serving \ac{BS} is located at a distance of $\tau_{\rm bu}$ to the typical user,} the corresponding \ac{EMFE} must not exceed the maximum allowable limit $\mathcal{W}_{\max}$ with high probability $\rho$~\cite{Joint}. 
For safety, the worst-case \ac{EMFE} from \acp{BS}, i.e., $\mathcal{W}_{\rm DL}$, should be considered in designing $\tau_{\rm bu}$. Mathematically,  
\begin{align}\label{eq:xcom}
    \tau_{\rm bu} \triangleq \inf_{t_{\rm bu}\in \mathbb{R}} \left\{t_{\rm bu}:   F_{\mathcal{W}_{\rm DL}|t_{\rm bu}}(\mathcal{W}_{\max}) \geq \rho  \right\}.
\end{align}
From \eqref{eq:inverseF}, $\tau_{\rm bu}$ satisfies
$F^{-1}_{\mathcal{W}_{\rm DL}|\tau_{\rm bu}}(\rho)=\mathcal{W}_{\max}$.
Clearly, access to the area centered around a \ac{BS} with a radius of $\tau_{\rm bu}$ is strictly prohibited as individuals within that area have a high possibility of experiencing \ac{EMFE} larger than $\mathcal{W}_{\max}$. Conversely, areas beyond the radius of $\tau_{\rm bu}$ are permissible as the \ac{EMFE} remains within the acceptable limit $\mathcal{W}_{\max}$ with a high probability.
 
It is worth considering whether a similar \ac{CD} should be set to account for the \ac{EMFE} from \ac{RIS} reflection. 
As shown in Sec.~\ref{subsec:DLEMF},  
if the typical user at $\mathbf{u}_0$ establishes a cascaded link with the \ac{BS} at $\mathbf{b}_0$ aided by the \ac{RIS} at $\mathbf{r}_0$, the \ac{RIS} can potentially lead to significant \ac{EMFE}. In this case, the aggregate downlink \ac{EMFE} is $\mathcal{W}_{\rm CL}$ in \eqref{eq:Wq}.
{Therefore, we define the \ac{CD} for the \ac{RIS}, denoted by $\tau_{\rm ru}$, as }
\begin{align}\label{eq:ycom_aver}
    \tau_{\rm ru} \triangleq \inf_{t_{\rm ru}\in \mathbb{R}} \left\{t_{\rm ru}:   F_{\mathcal{W}_{\rm CL}|t_{\rm ru}}(\mathcal{W}_{\max}) \geq \rho  \right\},
\end{align}
where $F_{\mathcal{W}_{\rm CL}|t_{\rm ru}}(\omega) = \mathbb{E}_{t_{\rm bu},t_{\rm br}} [ F_{\mathcal{W}_{\rm CL}|t_{\rm bu},t_{\rm ru},t_{\rm br}}(\omega) ]$.
Furthermore, due to the passive nature of the \ac{RIS} and its limited signal processing capability, the \ac{RIS} is usually controlled by the \ac{BS}~\cite{tutorialRIS}.  
As a result, the \ac{BS}-\ac{RIS} distance $t_{\rm br}$ is known in practical scenarios.  
Moreover, when $t_{\rm ru}$ and $t_{\rm br}$ are given, 
\begin{align}\label{eq:tbu}
    t_{\rm bu}=\sqrt{t_{\rm ru}^2+t_{\rm br}^2-2t_{\rm ru}t_{\rm br}\cos\vartheta_0}, 
\end{align} 
where {$\vartheta_0\sim\mathcal{U}(-\pi,\pi]$ } is the included angle from the RIS-user link to the RIS-BS link, {as shown in Fig.~\ref{fig:sys}.}
Therefore, we propose a metric for designing the conditional \ac{CD} of the \ac{RIS} as follows. 
\begin{align}\label{eq:ycom}
    \hat \tau_{\rm ru}(t_{\rm br}) \triangleq \!\!\inf_{t_{\rm ru}\in \mathbb{R}} \!\!\left\{t_{\rm ru}\!:\!   F_{\mathcal{W}_{\rm CL}|t_{\rm ru},t_{\rm br}}(\mathcal{W}_{\max}) \geq \rho  \right\},
\end{align}
where $F_{\mathcal{W}_{\rm CL}|t_{\rm ru},t_{\rm br}}(\omega) = \mathbb{E}_{\vartheta_0} [ F_{\mathcal{W}_{\rm CL}|\vartheta_0,t_{\rm ru},t_{\rm br}}(\omega) ]$, where  
$F_{\mathcal{W}_{\rm CL}|\vartheta_0,t_{\rm ru},t_{\rm br}}(\omega)$ is obtained by substituting \eqref{eq:tbu} into \eqref{eq:ccdf_q}.
Specifically, $\hat \tau_{\rm ru}$ is the minimum allowable separation between the RIS and the user conditioned on $t_{\rm br}$.
{For clarity, we refer to $\tau_{\rm ru}$ and $\hat \tau_{\rm ru}$ as {\em average and conditional \acp{CD}} of the RIS, respectively.} 
From \eqref{eq:inverseF}, $\tau_{\rm ru}$ and $\hat \tau_{\rm ru}$ satisfy $F^{-1}_{\mathcal{W}_{\rm CL}|\tau_{\rm ru}}(\rho)=\mathcal{W}_{\max}$ and $F^{-1}_{\mathcal{W}_{\rm CL}|\hat \tau_{\rm ru},t_{\rm br}}(\rho) = \mathcal{W}_{\max}$.

From the above discussion, $\tau_{\rm bu}$, $\hat \tau_{\rm ru}$, and $\tau_{\rm bu}$ can be identified as the optimal solutions of the optimization problems formulated in \eqref{eq:xcom}, \eqref{eq:ycom_aver}, and \eqref{eq:ycom}, respectively.  
These optimization problems can be solved by efficient one-dimension search algorithms such as bi-sectional~\cite{eiger1984bisection} or golden section methods~\cite{kiefer1953sequential}. To facilitate more efficient identification on \acp{CD},
we further provide approximate closed-form expressions for $\tau_{\rm bu}$, $\hat \tau_{\rm ru}$ and $\tau_{\rm bu}$. 
It is important to note that the exposure is primarily dominated by the received serving power density, especially when the serving link distances are short. Moreover, the CDs are typically quite short~\cite{Joint}. Based on these observations,
we ignore the exposure caused by the interfering signals,  i.e., $\mathcal{W}_2$, which allows us to derive approximate CDs as follows.
\begin{prop}[Compliance Distance]\label{prop:CD}
Considering that the \ac{EMFE} caused by the serving signals is dominant when the serving link distances are quite short, the BS CD, conditional RIS CD, and average RIS CD can be approximated as
\begin{align}\label{eq:CDbu}
            & \tau_{\rm bu} \approx  \bigg(\frac{ p_{\rm b} G_{\rm b} }{ 4 \pi \mathcal{W}_{\max}} F_{H_{\rm L}}^{-1}(\rho)\bigg)^{\frac{1}{\alpha_{\rm L}}},
\end{align}
\begin{align}\label{eq:CDru_tbr}
            \hat \tau_{\rm ru}(t_{\rm br}) \approx \bigg(\frac{ p_{\rm b} G_{\rm b} G_{\rm r} t_{\rm br}^{-\alpha_{\rm L}} } {4 \pi  \mathcal{W}_{\max}} F_{H_{\rm L}}^{-1}( \rho)
\bigg)^{\frac{1}{\alpha_{\rm L}}},
\end{align}
\begin{align}\label{eq:CDru}
            \tau_{\rm ru} \approx \bigg( \frac{ p_{\rm b} G_{\rm b} G_{\rm r}  }
{4 \pi \mathcal{W}_{\max}} F_{\alpha_{\rm L}}^{-1}(\rho) \bigg)^{\frac{1}{\alpha_{\rm L}}},
\end{align}
where $F_{H_{\rm L}}^{-1}(\rho)$ is the inverse function of $F_{H_{\rm L}}(h)$ given in \eqref{eq:CDF_ssfading},  $F_{\alpha_{\rm L}}^{-1}(\rho)$ is the inverse function of $F_{\alpha_{\rm L}}(h)$, and 
\begin{align}
            F_{\alpha_{\rm L}}(h)\!=\!\frac{\alpha_{\rm L}(m_{\rm L}h)^{m_{\rm L}}}{\Gamma(m_{\rm L})} \!\!\int_{0}^{\infty} \!\! 
 t^{m_{\rm L}\alpha_{\rm L}-1} e^{-m_{\rm L}h t^{\alpha_{\rm L}} -\pi\lambda_{\rm b}t^2 }  \mathrm{d} t.
        \end{align}
For a special case of $\alpha_{\rm L}=2$, 
\begin{align}\label{eq:CDru_alpha2}
 \tau_{\rm ru} \approx 
 \sqrt{ \frac{ \lambda_{\rm b}\rho^{\frac{1}{m_{\rm L}}}  p_{\rm b} G_{\rm b} G_{\rm r} }{4  \mathcal{W}_{\max}m_{\rm L}(1-\rho^{\frac{1}{m_{\rm L}}}) } }. 
\end{align}
\begin{IEEEproof}
See Appendix~\ref{app:CD}.
\end{IEEEproof}        
\end{prop}

Note that $F_{H_{\rm L}}^{-1}(\rho)$  can be evaluated by the built-in function in MATLAB, i.e., ${\rm gammaincinv}$.
In addition, $F_{\alpha_{\rm L}}(h)$ depends on the path loss exponent $\alpha_{\rm L}$ and the Nakagami-m shaping parameter $m_{\rm L}$.  Therefore, $F_{\alpha_{\rm L}}^{-1}(\rho)$ can be easily evaluated offline. The accuracy of \eqref{eq:CDbu}-\eqref{eq:CDru} will be verified in Sec.~\ref{sec:results}.
With Proposition~\ref{prop:CD}, we can efficiently determine CDs by using powerful math tools. Moreover, Proposition~\ref{prop:CD} provides intuitive insights into the relationship between the CDs and system parameters, offering valuable guidance for system design. For example, from \eqref{eq:CDbu}-\eqref{eq:CDru}, the CDs are proportional to the transmit power and the number of BS antennas (or the number of RIS elements) and are inversely proportional to the path loss exponent. For the conditional RIS CD in \eqref{eq:CDru_tbr}, 
a decrease in $t_{\rm br}$ necessitates an increase in the CD. This is because a shorter BS-RIS distance results in higher incident signal power at the RIS, requiring a larger CD to ensure that the \ac{EMFE} constraint is met.
Moreover, from \eqref{eq:CDru_alpha2},  
we observe that the average RIS CD is also influenced by the BS density $\lambda_{\rm b}$. Higher BS densities may lead to a shorter length of the cascaded link, thereby affecting $\tau_{\rm ru}$.

\subsection{Joint Downlink SINR and \ac{EMFE} Analysis}\label{subsec:Joint}
The following theorem derives the joint distribution of downlink SINR and \ac{EMFE} in  \eqref{eq:joint}.
\begin{theorem}[Downlink Joint Probability]\label{theorem:Joint}
The joint probability that $\Upsilon>\gamma$ and $\mathcal{W}\le \omega$ in \eqref{eq:joint} is 
\begin{align}
&J_{\Upsilon,\mathcal{W}}(\gamma,\omega)  = \mathbb{E}_{t_{\rm bu},t_{\rm ru},t_{\rm br}} \bigg[ \sum_{q\in\mathcal{S}} \mathcal{A}_{q} J_{{\Upsilon}_{q},\mathcal{W}_{q}|\mathcal{T}_{q}} (\gamma,\omega) \bigg],
\end{align}
where 
$J_{\Upsilon_q,\mathcal{W}_q|\mathcal{T}_q}(\gamma,\omega)$ is given by 

 \begin{align}\label{eq:joint_q}
&J_{\Upsilon_q,\mathcal{W}_q|\mathcal{T}_q}(\gamma,\omega)
=\mathbb{P}(\Upsilon_q>\gamma,\mathcal{W}_q\le \omega|\mathcal{T}_q)
=U(\gamma,\omega)\times
\nonumber\\&
\Bigg(
\frac{1}{2} F_{H_{q}}(h)\bigg|^{\frac{\mathcal{E} \omega}{ P^{\rm a}_{q} }}_{\frac{\sigma^2 \gamma }{ P^{\rm a}_{q} }}
-\int_{0}^{\infty}\frac{1}{\pi x}{\rm Im}\bigg[ 
\big(\Xi^{q}_1(-jx)\exp(jx\sigma^2)
\nonumber\\&\quad+\Xi^{q}_2(jx)\exp(-jx \mathcal{E} \omega )\big)
\mathcal{L}_{I_{\rm B}|t_{\rm bu}}(-jx) \bigg ] {\rm d} x 
\Bigg),
\end{align}
where $U(\gamma,\omega)=\mathbbm{1}(\mathcal{E} \omega - \sigma^2\gamma\ge 0)$ is an indicator function, $f(h)|^{H_2}_{H_1}=f(H_2)-f(H_1)$, $P^{\rm a}_q$ is given in \eqref{eq:Pr_dLN} and \eqref{eq:Pr_cL}, and 
\begin{subequations}\label{eq:Xi}
\begin{align}\label{eq:xi1}
& \Xi^{q}_1(-jx) = -\exp\left(-\frac{jx P^{\rm a}_{q} + m_{q} \gamma}{\gamma} h \right) 
\sum_{k=0}^{m_{q}-1}\frac{(m_{q}h)^k}{k!}\times
\nonumber\\&\left(1-\frac{jx P^{\rm a}_{q} }{\gamma (k+1)} 
\sum_{\iota=0}^{k}\frac{h^{-\iota}}{(\frac{jx P^{\rm a}_{q} +m_{q}\gamma}{\gamma})^{\iota+1}} A^{\iota+1}_{k+1}
\right)
\bigg|_{\frac{\sigma^2 \gamma }{ P^{\rm a}_{q} } }^{\frac{P_1}{ P^{\rm a}_{q} }},
\end{align}    
 \begin{align}\label{eq:xi2}
& \Xi^{q}_2(jx)=
-\exp\left(-(m_{q}-jx  P^{\rm a}_{q} ) h \right) 
\sum_{k=0}^{m_{q}-1}\frac{(m_{q}h)^k}{k!}\times
\nonumber\\&\left(1+\frac{jx  P^{\rm a}_{q} }{k+1} \sum_{\iota=0}^{k}\frac{h^{-\iota}}{(m_{q}-jx P^{\rm a}_{q})^{\iota+1}} A^{\iota+1}_{k+1}
\right)
\bigg|^{\frac{\mathcal{E} \omega}{ P^{\rm a}_{q} } }_{\frac{P_1}{ P^{\rm a}_{q} }},
\end{align}  
\end{subequations}
where {$m_q \in \mathbb{N}$}, $P_1=\frac{\gamma (\mathcal{E} \omega+\sigma^2)}{1+\gamma}$, and $A^{\iota+1}_{k+1}=\frac{(k+1)!}{(k-\iota)!}$.
\begin{IEEEproof}
See Appendix~\ref{app:lemma:JOINT_q}.
\end{IEEEproof}
\end{theorem}

\section{Uplink SINR and \ac{EMFE}}\label{sec:ULmodel}
This section focuses on the uplink communication from the users to the BSs.
We provide the SINR at the tagged BS and the \ac{EMFE} at the typical user.
We also define the
performance metrics to characterize the marginal and joint distributions of uplink SINR and \ac{EMFE}. 

\subsection{Uplink SINR}
In this subsection, we introduce the uplink transmit power at the typical user, the received power at the tagged BS, and the interference from other users.

\subsubsection{Power Control Mechanism}\label{subsec:PowerControl}

Generally, the user deploys a distance-dependent power control mechanism to compensate for the path loss~\cite{andrews2016primer}.
The transmit power of the typical user engaged in the direct LoS link is
\begin{equation}\label{eq:PtranUL_dL}
p_{\rm u,DL}(t_{\rm bu})=\begin{cases}
{p_{0}t_{\rm bu}^{\alpha_{\rm L}\epsilon}}, & \text{ if } t_{\rm bu}< T_{\max},\\
 p_{\rm max}, & {\rm otherwise},
\end{cases}
\end{equation}
where $p_{0}$ is a constant, $\epsilon\in(0,1]$ is the power control factor, $p_{\rm max}$ is the maximum transmit power of the user, and $T_{\max}=\left(p_{\rm max}/p_{0}\right)^{1/(\alpha_{\rm L}\epsilon)}$.
The transmit power of the typical user engaged in the direct NLoS link is 
\begin{equation}\label{eq:PtranUL_dN}
p_{\rm u,DN}(t_{\rm bu})=\begin{cases}
{p_{0}t_{\rm bu}^{\alpha_{\rm N}\epsilon}}, & \text{ if } t_{\rm bu}< (T_{\max})^{\frac{\alpha_{\rm L}}{\alpha_{\rm N}}},\\
 p_{\rm max}, & {\rm otherwise}.
\end{cases}
\end{equation}
With the aid of the RIS at $\mathbf{r}_0$, the path loss in the cascaded LoS link can be compensated by the transmit power and the RIS gain $G_{\rm r}$. Therefore, the uplink power control is not only dependent on the distance but also on the RIS gain, i.e., the transmit power of the typical user engaged in the cascaded LoS link is
\begin{equation}\label{eq:PtranUL_cL}
p_{\rm u,CL}(t_{\rm br},t_{\rm ru})=\begin{cases}
{p_{0}\big((t_{\rm br}t_{\rm ru})^{\alpha_{\rm L}}G_{\rm r}^{-1}\big)^\epsilon}, & \!\!\!\text{if } t_{\rm br}t_{\rm ru}< T_{\max} G_{\rm r}^{\frac{1}{\alpha_{\rm L}}},\\
 p_{\rm max}, & \!\!\!{\rm otherwise}.
\end{cases}
\end{equation}

\subsubsection{Received Serving Signal Power} 
Under the power control mechanism, the received serving signal power at the tagged BS via different types of serving links is 
\begin{subequations}
\begin{align}\label{eq:PrecUL_dL}
P'_{\rm DL}  & = p_{\rm u,DL}(t_{\rm bu}) G_{\rm b}\zeta t_{\rm bu}^{-\alpha_{\rm L}} H'_{{\rm DL}},
\\ \label{eq:PrecUL_cL}
P'_{\rm CL} &=p_{\rm u,CL}(t_{\rm br},t_{\rm ru})G_{\rm r}G_{\rm b}\zeta(t_{\rm ru}t_{\rm br})^{-\alpha_{\rm L}} H'_{{\rm CL}},
\\ \label{eq:PrecUL_dN}
P'_{\rm DN} & = p_{\rm u,DL}(t_{\rm bu}) G_{\rm b}\zeta t_{\rm bu}^{-\alpha_{\rm N}} H'_{{\rm DN}},
\end{align}
\end{subequations}
where $H'_{q}$, $q\in\mathcal{S}$, is the small-scaling fading coefficient. 

\subsubsection{Active Users and Uplink Interference}\label{subsec:ULinterference} 
Considering that only a single user is associated with a BS at each resource block, the number of active users is equal to the number of BSs. Therefore, the locations of active users are approximated by a \ac{PPP} $\Psi_{\rm ua}$ with density $\lambda_{\rm ua}=\lambda_{\rm b}$~\cite{andrews2016primer}.  
Moreover, an active user at $\mathbf{u}_i$ acts as an interferer to the tagged BS at $\mathbf{b}_0$ if it is not associated with the tagged BS, i.e., the tagged BS is not the nearest BS to this active user.
The probability of this interfering event is equivalent to the probability that there exists at least one BS that lies closer to $\mathbf{u}_i$ than the tagged BS. Based on the void probability of a PPP, this interfering probability is given by $1-\exp(-\pi\lambda_{\rm ua} {d'_i}^2)$, where $d'_i=||\mathbf{u}_i-\mathbf{b}_0||$.
Therefore, the locations of interfering users (i.e., all active users except the typical user) can be modeled by a non-homogeneous PPP $\Psi_{{\rm iu}}=\Psi_{\rm ua}\setminus\{\mathbf{u}_0\}$ with distance-dependent density $\lambda_{\rm iu}(d'_i)=\lambda_{\rm ua} (1-\exp(-\pi\lambda_{\rm ua} {d'_i}^2))=\lambda_{\rm b} (1-\exp(-\pi\lambda_{\rm b} {d'_i}^2))$  relative to the tagged BS~\cite{andrews2016primer,singh2015joint,qin2023unveiling}.
Furthermore, we divide $\Psi_{\rm iu}=\Psi_{\rm iu,L}\cup \Psi_{\rm iu,N}$, where the link between the interfering user at $\mathbf{u}_i\in\Psi_{{\rm iu}, v}$ and the tagged BS at $\mathbf{b}_0$ is LoS ($v={\rm L}$) or NLoS ($v={\rm N}$).
Moreover, different from the downlink case where an interfering BS has a constant transmit power, the transmit power of an interfering user depends on the type of its serving link, i.e., $q \in \mathcal{S}$.
Therefore, we have $\Psi_{{\rm iu}, v}=\sum_{q\in \mathcal{S}} \Psi_{{\rm iu},v,q}$.
For example, if an interfering user is located at $\mathbf{u}_i\in\Psi_{{\rm iu}, v, {\rm CL}}$, where $q={\rm CL}$, the transmit power of the user is $p_{\rm u,CL}(t_{{\rm br},i},t_{{\rm ru},i})$.
Under this model, the uplink interference at the tagged BS is
\begin{align}\label{eq:IU}
I_{\rm U}&=\sum_{v\in\{\rm L,N\}} \sum_{q\in\mathcal{S}} 
\sum_{i,\mathbf{u}_i\in \Psi_{{\rm iu},v,q} } \!\!\!\!p_{{\rm u},q} \times
 G_{\rm B}(\Delta'_i)\zeta {d'_i}^{-\alpha_v} H'_{v,i},
\end{align}
where ${d'_i}=||\mathbf{u}_i-\mathbf{b}_0||$, $H'_{v,i}$ is the small-scale fading coefficient between $\mathbf{u}_i$ and $\mathbf{b}_0$, $H'_{v,i}$ follows the distribution in \eqref{eq:ssfading}, and $G_{\rm B}(\Delta'_i)$ is the antenna gain of BS $\mathbf{b}_0$ at  $\Delta'_i$ (defined in Sec.~\ref{subsec:antenna}). 
\subsubsection{Uplink SINR and Coverage Probability}

The uplink \ac{SINR}, denoted by $\Upsilon'_q, q\in \mathcal{S}$, is
\begin{align}\label{eq:SINRq_UL}
{\Upsilon}'_{q}=\frac{P'_{q}}{\sigma'^2+I_{\rm U}}.
\end{align}
where $\sigma'^2$ is the noise power at a BS. 
We define the {\em uplink coverage probability} at a predefined threshold $\gamma'$ as
\begin{align}\label{eq:CCDF_UL}
\bar F_{\Upsilon'}(\gamma') &= \mathbb{P}\left \{ {{\Upsilon}'}>\gamma' \right \}=\sum_{q\in\mathcal{S}} \mathbb{P}\left \{ {{\Upsilon}'_q}>\gamma' \right \} \mathcal{A}_{q},
\end{align}
where $\mathcal{A}_{q}$ is the association probability given in Sec.~\ref{subsec:assoProb}.

\subsection{Uplink \ac{EMFE}}

Recall that the downlink exposure is induced by BSs and RISs (at distances in meters to the typical user), and the received power density can be used as a metric to measure the downlink exposure.
Difference from the downlink exposure, the uplink exposure is dominated by the typical user's personal mobile device (at a distance in centimeters to the typical user)~\cite{kuehn2019modelling,Joint,qin2023unveiling}.
Therefore, we use another metric, 
\ac{SAR}, to assess the uplink exposure. 
Specifically, \ac{SAR} is the power absorbed per mass of the exposed tissue, which relies on user-specific properties such as age (adult or child), usage (data or voice call), and posture (standing or sitting).
Mathematically, SAR is the product of the transmitted power of the device and the reference \ac{SAR} per transmit power~\cite{sar}.

Considering that the type of the uplink serving link is $q\in \mathcal{S}$, the corresponding uplink \ac{EMFE}, denoted by $\mathcal{W}'_{q} ~\left [ \rm W/kg \right ]$, can be quantified by SAR as~\cite{sar}
\begin{subequations}\label{eq:ULsar}
\begin{align}
\mathcal{W}'_{{\rm DL}} & = {\rm SAR}_{\rm ref} \times p_{\rm u,DL}(t_{\rm bu}),\\
\mathcal{W}'_{{\rm CL}} & = {\rm SAR}_{\rm ref} \times p_{\rm u,CL}(t_{\rm ru},t_{\rm br}),\\
\mathcal{W}'_{{\rm DN}} & = {\rm SAR}_{\rm ref} \times p_{\rm u,DN}(t_{\rm bu}),
\end{align}
\end{subequations}
where ${\rm SAR}_{\rm ref}\, \left[\rm \frac{W}{kg}/W\right]$ is the reference \ac{SAR}.
We define the {\em uplink compliance probability} at a constraint level $\omega'$ as
\begin{align}\label{eq:CDF_W_UL}
F_{\mathcal{W}'}(\omega') & = \mathbb{P}({\mathcal{W}'}\le \omega')= \sum_{q\in\mathcal{S}}  \mathbb{P}({\mathcal{W}'_{q}}\le \omega')\mathcal{A}_{q}.
\end{align}

\subsection{Joint Metric of Uplink SINR and \ac{EMFE}}
From Definition~\ref{def:joint}, we define the joint metric on the uplink SINR and \ac{EMFE} as
\begin{align}\label{eq:joint_UL}
\begin{split}
J_{\Upsilon',\mathcal{W}'}(\gamma',\omega')
&=\mathbb{P}(\Upsilon'>\gamma',\mathcal{W}'\le \omega') \\&
=\sum_{q\in\mathcal{S}}
\mathbb{P}(\Upsilon'_q>\gamma,\mathcal{W}'_q\le \omega')\mathcal{A}_q.     
\end{split}
\end{align} 

\section{Uplink Performance Analysis}\label{sec:ULanaly}
This section provides the expressions of the uplink performance metrics. 
Since the association rules are the same in both uplink and downlink cases, we can apply the association probabilities and the serving distance distributions in Sec.~\ref{subsec:assoProb} and Sec.~\ref{subsec:dist} to the uplink analysis.
We first provide the Laplace transform of the uplink interference for deriving the coverage probability. Then, we derive the uplink compliance probability and the first moment of uplink \ac{EMFE}, followed by the joint analysis of SINR and \ac{EMFE}.
\subsection{Laplace Transform of Uplink Interference}\label{subsec:UL_L_SINR}
 
As shown in Sec.~\ref{subsec:cov}, the Laplace transform of interference is a fundamental step to derive the coverage probability. This subsection characterizes the uplink interference $I_{\rm U}$ in \eqref{eq:IU} by its Laplace transform, which, however, is complicated due to the dynamic transmit power. For analytical tractability, we compute the average transmit power of an interfering user, which is used to simplify the Laplace transform of $I_{\rm U}$.  
Specifically, the transmit power of an interfering user is related to the type and distance of the serving link between the user and its associated BS.
From Slivnyak's theorem~\cite{haenggi2012stochastic}, 
the distributions of  $t_{{\rm bu},i}$, $t_{{\rm ru},i}$, or $t_{{\rm br},i}$ are the same as $t_{\rm bu}$, $t_{\rm ru}$, or $t_{\rm br}$, respectively.
Hence, we can compute the average transmit power of an interfering user as
\begin{align}
\bar p_{\rm u} &=  \mathbb{E}_{t_{{\rm bu},i},t_{{\rm ru},i},t_{{\rm br},i} } \bigg[ \sum_{q\in\mathcal{S}} \mathcal{A}_{q} p_{{\rm u},q} \bigg].   
\end{align}
Then, the uplink interference in \eqref{eq:IU} can be approximated as $I_{\rm U}=I_{\rm u,L}+I_{\rm u,N}$, where
\begin{align}\label{eq:approxIu}
I_{{\rm u},v} \approx
\sum_{i,\mathbf{u}_i\in{\Psi_{{\rm iu},v}}}{\bar p_{\rm u} G_{\rm b}(\Delta'_i)\zeta {d'_i}^{-\alpha_v} H'_{v,i}}, v\in\{\rm L,N\}.
\end{align}

\begin{lemma}\label{lemma:LT_UL_interference}
The Laplace transform of $I_{\rm U}$ is 
$\mathcal{L}_{I_{\rm U}}(s)=\mathcal{L}_{I_{\rm u,L}}(s)\mathcal{L}_{I_{\rm u,N}}(s)$, where $\mathcal{L}_{I_{{\rm u},v}}(s)$ is given by
\begin{align}
\begin{split}
&\mathcal{L}_{I_{{\rm u},v}}(s)=\mathbb{E}\left[\exp\left(-sI_{{\rm u},v}\right)\right] \approx\exp\bigg(\!\!\!
-\!2\pi \times
\\&\!\!\int_{0}^{\infty} \!\!\!\!
[1-\kappa_v(s \bar p_{\rm u}  \zeta {d'_i} ^{-\alpha_{v}})]
d'_i \lambda_{\rm iu}(d'_i) \mathcal{P}_{v}(d'_i)
\mathrm{d} d'_i
\bigg),
\end{split}
\end{align}
where $\kappa_v(\cdot)$ is given in Lemma~\ref{lemma:L_SINR} and $\lambda_{\rm iu}(d'_i)=\lambda_{\rm b} (1-\exp(-\pi\lambda_{\rm b} {d'_i}^2))$ is given in Sec.~\ref{subsec:ULinterference}. 
\begin{proof} 
By approximating the dynamic transmit power of an interfering user as a constant $\bar p_u$, we can derive the Laplace transform of uplink interference by using the methods in Appendix~\ref{app:lemma:L_SINR}. The accuracy of this approximation will be validated in Sec.~\ref{sub:ULresults}.
\end{proof}
\end{lemma}

\subsection{Uplink SINR Analysis}\label{subsec:ULcov}
With the approximated Laplace transform of uplink interference, we can derive the uplink coverage probability as follows.
\begin{theorem}[Uplink Coverage Probability]\label{theo:UL_sinr}
The \ac{CCDF} of the uplink SINR $\Upsilon'$ in \eqref{eq:CCDF_UL} is 
\begin{align}
\bar F_{\Upsilon'}(\gamma') = \mathbb{E}_{t_{\rm bu},t_{\rm ru},t_{\rm br}} \bigg[ \sum_{q\in\mathcal{S}}  \mathcal{A}_{q} \bar F_{{\Upsilon}'_{q}|\mathcal{T}_q} (\gamma') \bigg].
\end{align}
where 
\begin{align}\label{eq:ccdf_q_UL}
&\bar F_{{\Upsilon}'_{q}|\mathcal{T}_q}(\gamma')=\mathbb{P}({\Upsilon}'_{q}>\gamma'|\mathcal{T}_q)
\nonumber\\&\overset{(a)}{\approx}\sum_{k=1}^{m_{q}}\binom{m_{q}}{k}\left ( -1 \right )^{k+1} \exp(-k \beta_{q} s'_{q}\sigma'^2) \mathcal{L}_{I_{\rm U}}\left ( k \beta_{q} s'_{q} \right ) ,
\end{align} 
where 
(a) is from the methods introduced in the derivation of \eqref{eq:ccdf_q}, {$m_q \in \mathbb{N}$}, $s'_{\rm DL}= {m_{\rm L}\gamma'}\big({p_{\rm u,DL}(t_{\rm bu}) G_{\rm b}\zeta t_{\rm bu}^{-\alpha_{\rm L}}}\big)^{-1}$, $s'_{\rm CL}= {m_{\rm L}\gamma'}\big({p_{\rm u,CL}(t_{\rm ru},t_{\rm br}) G_{\rm b}\zeta t_{\rm ru}^{-\alpha_{\rm L}} t_{\rm br}^{-\alpha_{\rm L}} }\big)^{-1}$, and $s'_{\rm DN}=\ {m_{\rm N}\gamma'}\big({p_{\rm u,DN}(t_{\rm bu}) G_{\rm b}\zeta t_{\rm bu}^{-\alpha_{\rm N}}}\big)^{-1}$.
\end{theorem}

\subsection{Uplink \ac{EMFE} Analysis}\label{subsec:ULEMF}

Note that the uplink \ac{EMFE} in \eqref{eq:ULsar} is the product of the reference SAR and the transmit power, which is a function of $t_{\rm bu}$, $t_{\rm ru}$, or/and $t_{\rm br}$. Conditioned on serving link distances, the uplink \ac{EMFE} is a constant.
Hence, the \ac{CDF} of $\mathcal{W}'_{q}$ conditioned on $\mathcal{T}_q$, denoted by $F_{\mathcal{W}'_{q}|\mathcal{T}_q}(\omega')=\mathbb{P}(\mathcal{W}'_{q}\leq \omega'|\mathcal{T}_q)$, $q\in\mathcal{S}$, is
\begin{align} \label{eq:F_Wq_UL}
F_{\mathcal{W}'_{q}|\mathcal{T}_q}(\omega')&=
\mathbbm{1}(\mathcal{W}'_{q} \le \omega').
\end{align}  

 \begin{theorem}[Uplink \ac{EMFE}]\label{theo:UL_emf}
 The CDF of the uplink \ac{EMFE} $\mathcal{W}'$ in \eqref{eq:CDF_W_UL} is 
 \begin{align}\label{eq:emf_UL}
F_{\mathcal{W}'}(\omega')= \mathbb{E}_{t_{\rm bu},t_{\rm ru},t_{\rm br}} \bigg[ \sum_{q\in\mathcal{S}}  \mathcal{A}_{q} F_{\mathcal{W}'_{q}|\mathcal{T}_q} (\omega') \bigg].
\end{align}
The first moment of $\mathcal{W}'$ is $\mathbb{E}[\mathcal{W}'] =\mathbb{E}_{t_{\rm bu},t_{\rm ru},t_{\rm br}}\bigg[ \!\sum_{q\in\mathcal{S}} \!{\mathcal{W}'_{q}} \mathcal{A}_{q}\bigg]$.
 \end{theorem}

\subsection{Joint Uplink SINR and \ac{EMFE} Analysis}\label{subsec:ULJoint}
The joint analysis of the SINR and \ac{EMFE} in the uplink is given as follows.
\begin{theorem}[Uplink Joint Probability]\label{theorem:JOINT_UL}
The joint probability that $\Upsilon'>\gamma'$ and $\mathcal{W}'\le \omega'$ is 
\begin{align}
&J_{\Upsilon',\mathcal{W}'}(\gamma',\omega')  \!= \!\mathbb{E}_{t_{\rm bu},t_{\rm ru},t_{\rm br}} \bigg[ \sum_{q\in\mathcal{S}} \mathcal{A}_{q} J_{{\Upsilon}'_{q},\mathcal{W}'_{q}|\mathcal{T}_{q}} (\gamma',\omega') \bigg],\!  
\end{align}
where $J_{\Upsilon'_{q},\mathcal{W}'_{q}|\mathcal{T}_q}(\gamma',\omega')
=\bar F_{{\Upsilon}'_{q}|\mathcal{T}_q}(\gamma') F_{\mathcal{W}'_{q}|\mathcal{T}_q}(\omega')$.
\vspace{1mm}

\begin{IEEEproof}
The proof can be completed by following the similar steps in \eqref{eq:joint_1} in Appendix~\ref{app:lemma:JOINT_q} and thus is omitted here.
\end{IEEEproof}
\end{theorem}

\begin{table}[t!]\caption{Default values of system parameters.}
\vspace{-3mm}
\centering
\begin{center}
{\linespread{0.95}
\renewcommand{\arraystretch}{1.2} 
    \begin{tabular}{ | {c} | {c} || {c} | {c} | }
    \hline
        \hline
    {Parameters} & {Values} & {Parameters} & {Values} \\ \hline
    
    $\lambda_{\rm b}$ & $10^{-5} \rm{/m^2}$ & $\lambda_{\rm u}$ & $20\!\times\! 10^{-5} \rm{/m^2}$ \\ \hline
    $\lambda_{\rm o}$ & $50\!\times\! 10^{-5} \rm{/m^2}$ & $L_{\rm o}$ & $15~\rm m$   \\ \hline
    $(\alpha_{\rm L},\alpha_{\rm N})$~\cite{3gpp2017further} & $(2.09, 3.75)$~\cite{para0} &  $(m_{\rm L},m_{\rm N})$ & $(3, 1)$  
     \\ \hline 
    $(N_{\rm b},N_{\rm r})$ & $(8,16)$ &  
    $f$ & $28~\rm GHz$ \\ \hline
     {$\delta$} &  $1/\sqrt{2}$   & $(\mu,\epsilon)$ & $(0.12, 0.6)$\\ \hline
    $p_{\rm b}$ & $10 ~\rm W$ & $t_{\rm bu}$ & $100 ~\rm m$ \\ \hline
    $\sigma^2$ \cite{para1} & $8\times 10^{-12}~\rm W$ & $\sigma'^2$ \cite{para1} & $8\times 10^{-13} ~\rm W$ \\ \hline
    $(p_{0},p_{\max})$ \cite{Joint}& $(0.008, 200) ~{\rm mW}$ & $\rho$& $0.95$  \\ \hline
    ${\rm SAR}_{\rm ref} $ \cite{sar} &  $0.0053~\rm \frac{W}{kg}/W$  & $\mathcal{W}_{\max}$ & $10~\rm W/m^2$  \\ \hline
    \end{tabular}}
\end{center}
\label{tab:simulation}
\vspace{-2mm}
\end{table}

\section{Results and Discussions}\label{sec:results}
This section presents numerical results of the expressions derived throughout the paper and provides answers to \textbf{Q1-Q4} (posed in Sec. \ref{sec:intro}) based on numerical findings.
Moreover, we perform Monte Carlo simulations using the actual antenna pattern in \eqref{eq:Gact}.
The default values of system parameters are summarized in Table~\ref{tab:simulation}, unless otherwise specified. 
The noise power is calculated by $-174 ~{\rm dBm/Hz}+ 10\log_{10} BW + 10~ {\rm dB}$~\cite{para1}, where we consider bandwidth $BW=200~\rm MHz$ in downlink and $BW=20~\rm MHz$ in uplink.
The distance between the typical user and its serving BS is fixed at a default value of $t_{\rm bu}=100~\rm m$.\footnote{The analytical results for random $t_{\rm bu}$ can be obtained based on $\eqref{eq:fX}$.}

\subsection{Downlink Performance Evaluation}\label{sub:DLresults}
\subsubsection{Marginal Probability}

\begin{figure*}[t!]
\begin{minipage}{.325\textwidth}
 \centering
  \vspace{-3mm}
		\includegraphics[width=1\columnwidth]{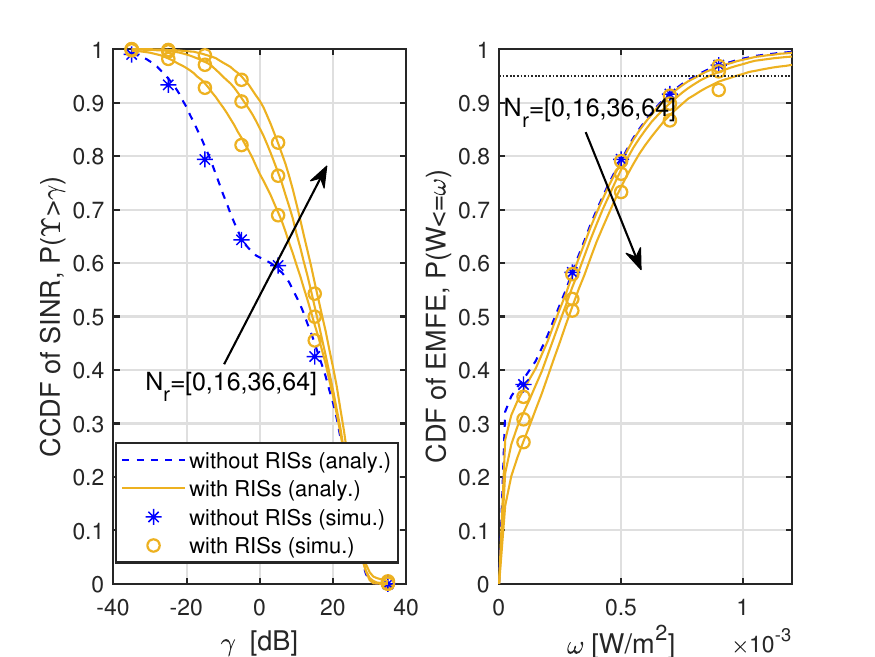} 
  \caption{Marginal distributions of downlink SINR ($\Upsilon$) and \ac{EMFE} ($\mathcal{W}$) for different values of the number of RIS elements ($N_{\rm r}$): coverage and compliance probabilities at $t_{\rm bu}=100~\rm m$.}
  \label{fig:marginal_DL}
\end{minipage}
\hfill
\begin{minipage}{.325\textwidth}
 		\centering
  \vspace{-3mm}
		\includegraphics[width=1\columnwidth]{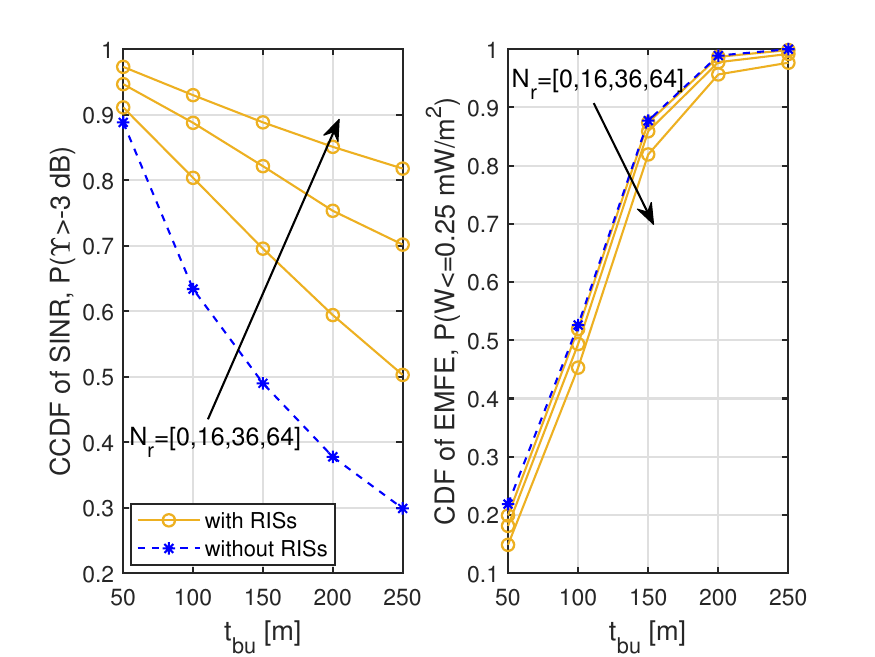}
  \caption{{Impact of the BS-user distance ($t_{\rm bu}$) 
 on the marginal distributions of downlink SINR ($\Upsilon$) and \ac{EMFE} ($\mathcal{W}$) at $\gamma=- 3~\rm dB$ and $\omega=0.25 ~\rm mW/m^2$.}}
  \label{fig:DL_distance}
    \end{minipage}
\hfill
\begin{minipage}{.325\textwidth}
		\centering
  \vspace{-3mm}
		\includegraphics[width=1\columnwidth]{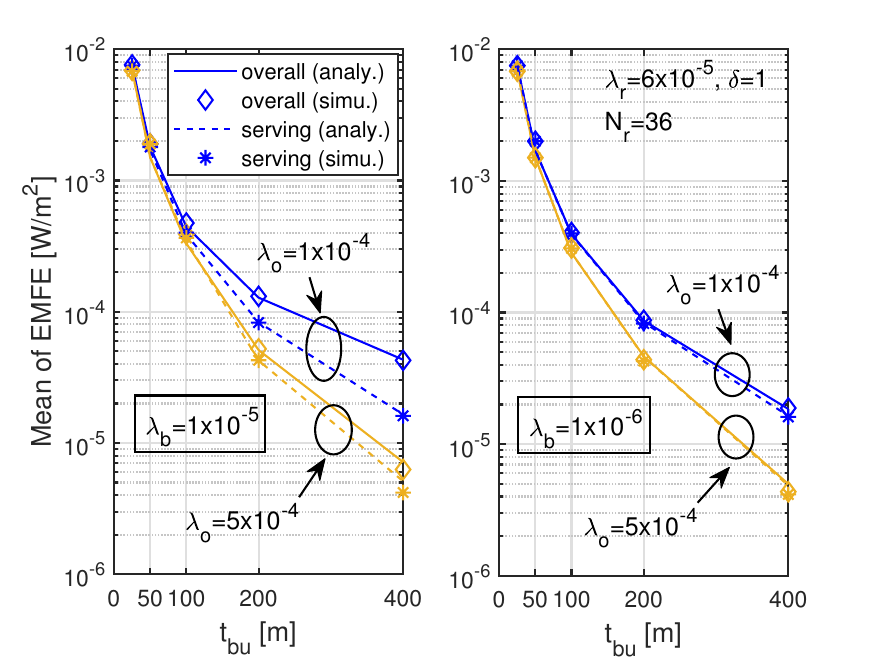}
  \caption{{Mean of downlink \ac{EMFE} caused by the serving signals and by the overall (serving and interfering) signals at different BS densities ($\lambda_{\rm b}$) and obstacle densities ($\lambda_{\rm o}$).}} 
  \label{fig:MeanEMF} 
    \end{minipage}
\end{figure*}

\begin{figure*}[t!]
\begin{minipage}{.325\textwidth}
\vspace{-3mm}
\centering
\includegraphics[width=0.95\columnwidth]{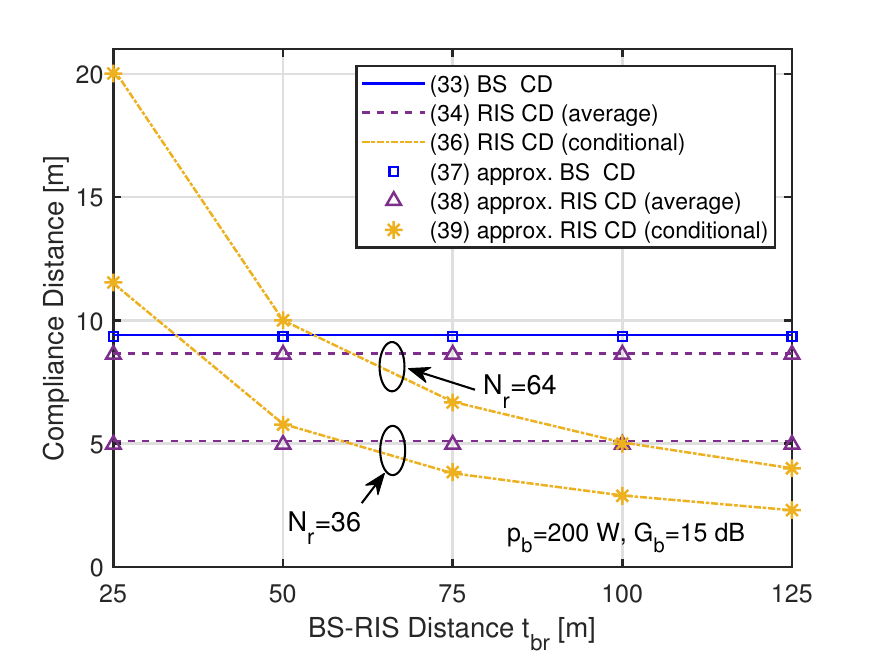}
\caption{The conditional RIS \ac{CD} $\hat \tau_{\rm ru} (t_{\rm br})$, the average RIS \ac{CD} $\tau_{\rm ru}$, and the BS \ac{CD} $\tau_{\rm bu}$ based on \eqref{eq:xcom}-\eqref{eq:ycom} and \eqref{eq:CDbu}-\eqref{eq:CDru}.}  
\label{fig:CD}
\end{minipage}
\hfill
\begin{minipage}{.325\textwidth}
  \vspace{-3mm}
		\centering
		\includegraphics[width=0.95\columnwidth]{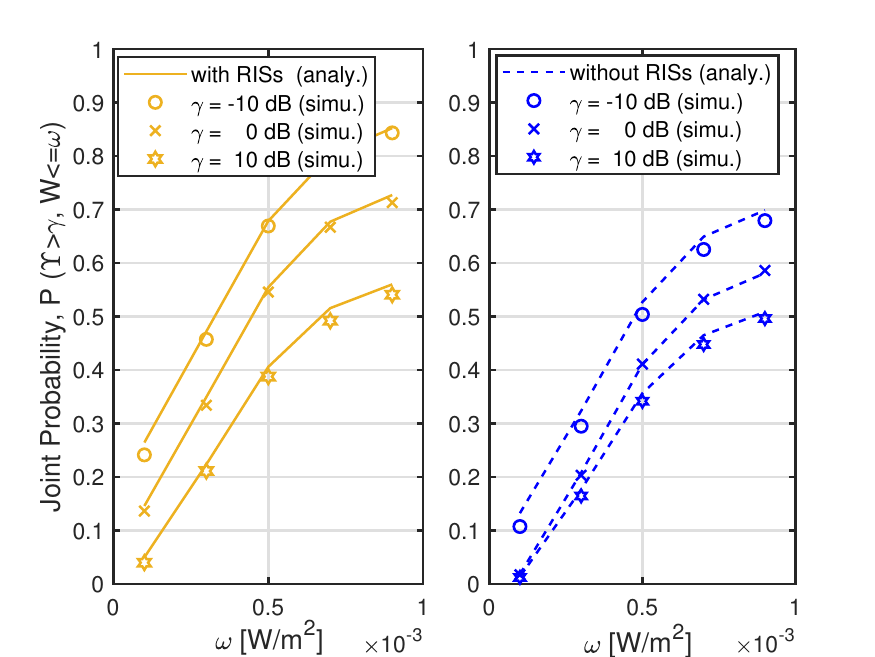}
  \caption{Joint distribution of downlink SINR and \ac{EMFE} for different values of $\gamma$ and $\omega$ with RISs (left) and without RISs (right).}
  \label{fig:DL_joint}
    \end{minipage}
\hfill
\begin{minipage}{.325\textwidth}
  \vspace{-3mm}
		\centering
		\includegraphics[width=0.95\columnwidth]{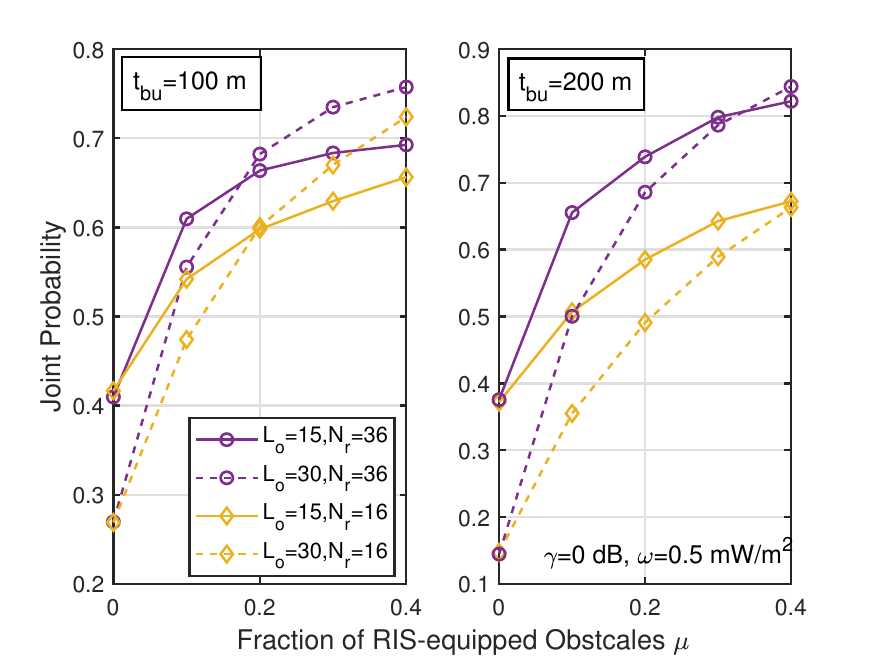}
  \caption{Joint metric on downlink SINR and \ac{EMFE} for different values of $\mu$, $N_{\rm r}$, and $L_{\rm o}$ at $t_{\rm bu}=100~\rm m$ and  $t_{\rm bu}=200~\rm m$.}  
  \label{fig:DL_Ratio}
    \end{minipage}
\end{figure*}


Fig.~\ref{fig:marginal_DL} plots the CCDF of SINR (coverage probability) and the CDF of \ac{EMFE} (compliance probability) in downlink for different values of the number of \ac{RIS} elements ($N_{\rm r}$), where the curve of $N_{\rm r}=0$ refers to the network without \acp{RIS}.
The simulation results closely match the analytical results, which validates Theorems~\ref{theorem:cov} and~\ref{theorem:emf}. 
Moreover, we see that deploying \acp{RIS} significantly improves coverage performance, while slightly degrading compliance performance.
For example, compared with the network without RISs, deploying RISs with $N_{\rm r}=64$ has roughly $30\%$ coverage enhancement at $\gamma=-5~\rm dB$ and increases the 95-th percentile of \ac{EMFE} from $0.8~\rm mW/m^2$ to $1~\rm mW/m^2$. This provides a clear answer to \textbf{Q1} from the downlink perspective, indicating that deploying RISs would exacerbate \ac{EMFE}.
The reason is that in the \ac{RIS}-assisted network, when the direct link between the typical user and its serving \ac{BS} is \ac{NLoS}, \acp{RIS} can provide a LoS link with less attenuation than the NLoS link.
Moreover, larger $N_{\rm r}$ provides higher \ac{RIS} gain ($G_{\rm r}$) to compensate for the path loss in the cascaded (BS-RIS-user) link.
This results in an increased received serving signal power and power density, thereby improving SINR at the expense of exacerbating \ac{EMFE}.

Fig.~\ref{fig:DL_distance} shows impact of the BS-user distance ($t_{\rm bu}$) on the coverage probability and the compliance probability at $\gamma=- 3~\rm dB$ and $\omega=0.25 ~\rm mW/m^2$ in downlink. The coverage probability gradually decreases with the increase of $t_{\rm bu}$ due to the increased path loss and NLoS probability of the serving link. For example, in the networks without RISs, the coverage probability is approximately $ 0.89 $ when $ t_{\rm bu} = 50~\rm m $, whereas it drops to $ 0.3 $ at $ t_{\rm bu} = 250~\rm m $.  
The small coverage range is a well-known challenge at millimeter wave (mmWave) communications ($f=28~\rm GHz$)~\cite{Millimeter}.  
Interestingly, exploiting RIS technologies can extend the coverage range. For example, to achieve a coverage probability greater than $0.8$ at $ \gamma=-3~\rm dB$, $t_{\rm bu}$ should be less than roughly $75~\rm m$ at the network without RISs; whereas $t_{\rm bu}$ can increase to $100~\rm m$ at the RIS-assisted network with $N_{\rm r}=16$, and even to $250~\rm m$ at the network with RISs with $N_{\rm r}=64$. Moreover, we also see that when the typical user is closer to its serving BS, i.e., decreasing $t_{\rm bu}$, the compliance probability decreases in both networks with and without RISs. Therefore, to ensure the compliance probability is high enough for safety constraints, the minimum value of $t_{\rm bu}$ is worth consideration, which will be shown in the next subsection. 

Fig.~\ref{fig:MeanEMF} presents mean values of downlink \ac{EMFE} caused by serving signals and by overall (serving and interfering) signals,  respectively. The close match between simulation and analytical results validates \eqref{eq:E_W} in Theorem~\ref{theorem:emf}. From the left sub-figure, we see that the gap between \ac{EMFE} caused by the serving signals and that caused by the overall signals gradually diminishes as the BS-user distance $t_{\rm bu}$ shortens. 
This indicates that the received serving power density increasingly dominates \ac{EMFE} at shorter distances, thereby justifying the reasonability of the approximation in Proposition~\ref{prop:CD}. 
We observe that a higher obstacle density tends to reduce the overall \ac{EMFE} since more interfering links are blocked. 
Additionally, a comparison between the left and right sub-figures reveals that a lower BS density also mitigates the overall \ac{EMFE} by reducing the number of radiating sources. 
These findings collectively emphasize the dominant role of the received serving power density in \ac{EMFE} in environments with short serving links, dense obstacles, or sparse BSs. 

\subsubsection{Compliance Distance (CD)}

Fig.~\ref{fig:CD} presents the conditional RIS \ac{CD} $\hat \tau_{\rm ru}$ (conditioned on BS-RIS distance $t_{\rm br}$), the average RIS \ac{CD} $\tau_{\rm ru}$, and the BS \ac{CD} $\tau_{\rm bu}$, under the conservative setting, where we consider the maximum transmit power ($200~\rm W$) and 
{the maximum antenna gain ($15~\rm dB$) at \acp{BS}}~\cite{Joint}.  
Specifically, $\hat \tau_{\rm ru}$, $\tau_{\rm ru}$, and $\tau_{\rm bu}$ are the minimum distances between a RIS/BS and a user to ensure that the 95-th percentile of \ac{EMFE} remains within the safe limit $\mathcal{W}_{\max}$. 
Comparing the CDs based on \eqref{eq:xcom}-\eqref{eq:ycom} by the bi-sectional method with that based on \eqref{eq:CDbu}-\eqref{eq:CDru} in Proposition~\ref{prop:CD}, we verify the accuracy of the approximate expressions for CDs.
Moreover, consistent with the observations from Proposition~\ref{prop:CD}, we see that the conditional and average RIS \acp{CD} increases with $N_{\rm r}$. This is because increasing $N_{\rm r}$ exacerbates the \ac{EMFE} level (as shown in Fig.~\ref{fig:marginal_DL}), requiring a larger fence (i.e., a longer CD) around the RIS.
{Particularly, at $N_{\rm r}=64$, the average RIS CD is $\tau_{\rm ru} \approx 8~\rm m$, which is comparable to the BS CD $\tau_{\rm bu}\approx 9.5 ~\rm m $. This implies that establishing a specific value for the RIS \ac{CD} is as necessary as setting the BS \ac{CD}, thereby answering \textbf{Q2}.}
We also see that the conditional \ac{CD} $\hat \tau_{\rm ru}$ is greater than the average \ac{CD} $\tau_{\rm ru}$ at small values of BS-RIS distance $t_{\rm br}$; whereas $\hat \tau_{\rm ru}<\tau_{\rm ru}$ at large values of $t_{\rm br}$.  
Intuitively, when the RIS is closer to the BS, the path loss in the BS-RIS link significantly decreases, resulting in stronger incident and reflected power at the RIS. 
Therefore, the fence around the RIS should be enlarged to prevent users from entering areas with excessive \ac{EMFE}.

\subsubsection{Joint Probability}

Fig.~\ref{fig:DL_joint} shows the joint distribution of the downlink SINR ($\Upsilon$) and \ac{EMFE} ($\mathcal{W}$), i.e., $\mathcal{J}_{\Upsilon,\mathcal{W}}(\gamma,\omega)$. 
We can see that the simulation results closely match the analytical results, validating Theorem~\ref{theorem:Joint}.
Compared with the network without RISs,
the RIS-assisted network has a higher probability of achieving a given SINR threshold $\gamma$ and adhering to a specified  \ac{EMFE} constraint level $\omega$ simultaneously.   
For example, under the constraint of \ac{SINR} at $\gamma=0~\rm dB$, the probability that \ac{EMFE} is less than $\omega=0.5~\rm mW/m^2$ is up to $0.55$ in the network with RISs, while it is only $0.4$ in the network without RISs. 
From Fig.~\ref{fig:marginal_DL} and Fig.~\ref{fig:DL_joint}, we can conclude that deploying RIS would improve SINR at the expense of exacerbating \ac{EMFE} in downlink; whereas the benefits of deploying RISs in enhancing coverage performance outweigh the compromises in \ac{EMFE} compliance performance. This answers \textbf{Q3} from the downlink perspective.

To investigate which deployment strategy is better for the downlink performance (\textbf{Q4}),
Fig.~\ref{fig:DL_Ratio} shows the impact of the fraction of RIS-equipped obstacles $\mu$, the number of RIS elements $N_{\rm r}$, and the average length of obstacles $L_{\rm o}$ on the downlink performance at $\gamma=0\,\rm dB$ and $\omega=0.5\,\rm mW/m^2$.
A larger $\mu$, i.e., a higher \ac{RIS} density, increases the existence probability of the cascade LoS link in \eqref{eq:existence} and reduces the length of the cascade LoS link. 
Consequently, the received serving signal power and power density strengthen, positively impacting the SINR while negatively impacting the \ac{EMFE}.  
We see that increasing $\mu$ improves system performance at $L_{\rm o}=15\,\rm m$ and $N_{\rm r}=16$, implying that the positive effect of increasing $\mu$ in coverage dominates. This can be considered as a potential downlink deployment strategy for \textbf{Q4}.
Interestingly, there is a different trend at low and high fractions of RIS-equipped obstacles as $L_{\rm o}$ increases. 
This is because the increased $L_{\rm o}$ not only decreases the LoS probability in the serving link but also in the interfering link, resulting in a decrease in both received power and the received power density in the serving link and the interfering link.
The lowered received power density and reduced interference positively impact \ac{EMFE} and SINR, while the decreased received power in the serving link negatively impacts SINR. For example, for $t_{\rm bu}=100~\rm m$, when $\mu\in [0,0.17)$, the detrimental impact of increasing $L_{\rm o}$ dominates, leading to performance degradation; contrarily, when $\mu\in (0.2,0.4]$, the density of RISs is sufficient to provide cascaded LoS links,  compensating for the decreased LoS probability in the serving link. Thus, increasing $L_{\rm o}$ results in enhanced performance. 
These observations reveal the interdependence of the RIS deployment and the obstacle environment on the SINR and \ac{EMFE}, providing valuable insights into RIS configuration to balance the trade-off between coverage and compliance performance.

\begin{figure*}[t!]

\begin{minipage}{.325\textwidth}
\centering
  \vspace{-3mm}
		\includegraphics[width=0.95\columnwidth]{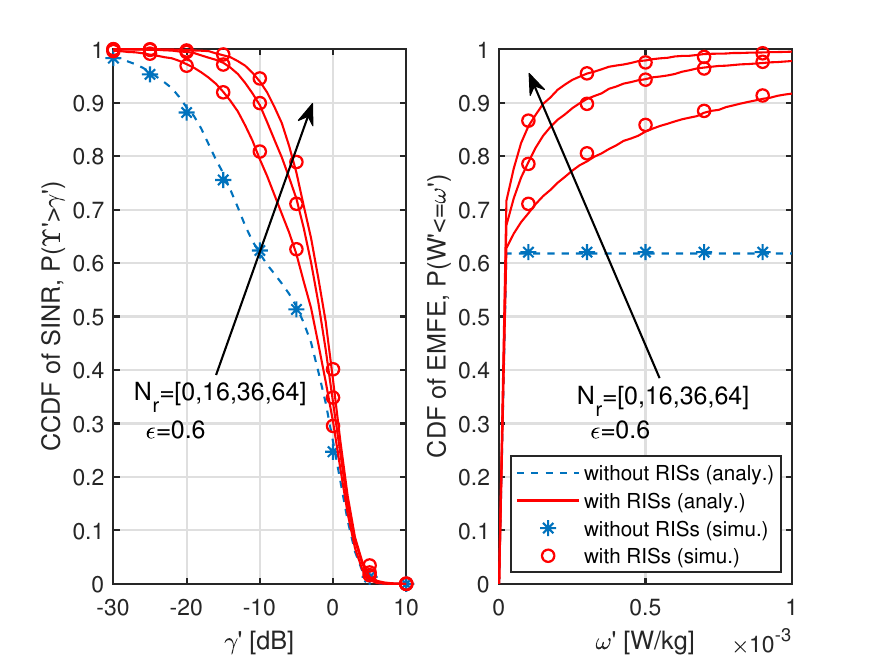}
  \caption{Marginal distributions of uplink SINR ($\Upsilon'$) and \ac{EMFE} ($\mathcal{W}'$).}
  \label{fig:marginal_UL}
\end{minipage}
    \hfill
\begin{minipage}{.325\textwidth}
  \vspace{-3mm}
		\centering
\includegraphics[width=0.95\columnwidth]{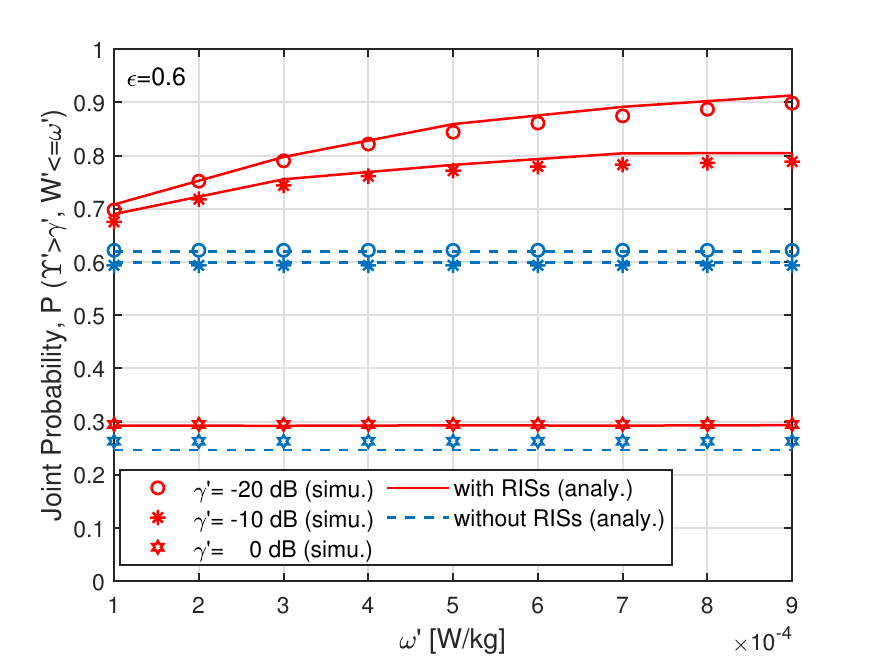}
  \caption{Joint distribution of uplink SINR and \ac{EMFE} for different values of $\gamma'$ and $\omega'$.}
  \label{fig:UL_joint}
\end{minipage}
\hfill
\begin{minipage}{.325\textwidth}
  \vspace{-3mm}
		\centering
\includegraphics[width=0.95\columnwidth]{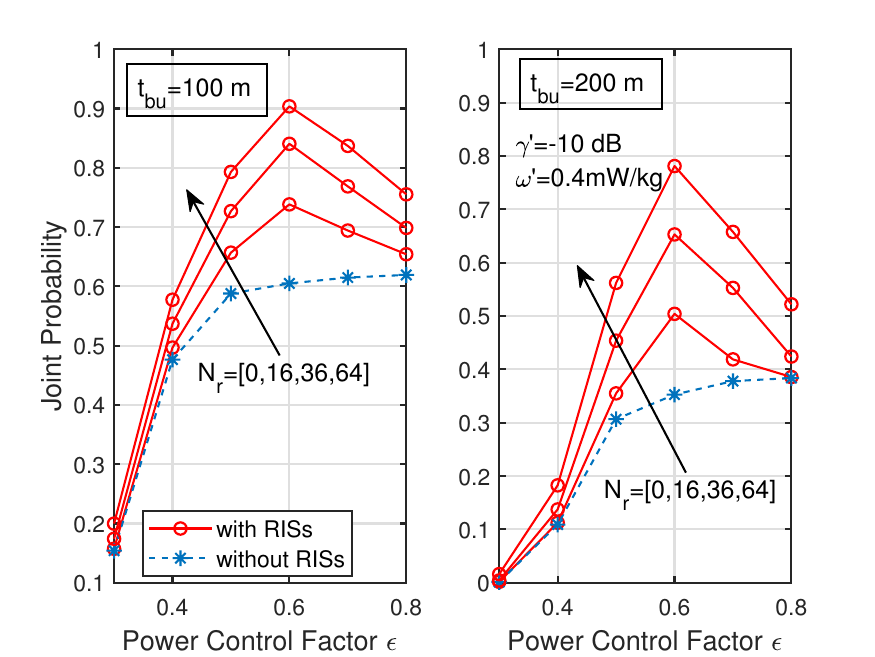}
  \caption{Joint metric on uplink SINR and \ac{EMFE} for different power control factors $\epsilon$.}
  \label{fig:UL_epsilon}
    \end{minipage}
\end{figure*}

\subsection{Uplink Performance Evaluation}\label{sub:ULresults}

\subsubsection{Marginal Probability}

Fig.~\ref{fig:marginal_UL} plots the CCDF of SINR (coverage probability) and the CDF of \ac{EMFE} (compliance probability) in uplink for different values of $N_{\rm r}$.
The simulation results closely match the analytical results, which validates Theorems~\ref{theo:UL_sinr} and~\ref{theo:UL_emf}. Additionally, it verifies the applicability of Lemma~\ref{lemma:LT_UL_interference} in approximating the uplink interference. 
\textcolor{black}{We see that in the network without RISs, the compliance probability is constant within $\omega'\in [0.1,1] ~\rm mW/kg$. This is because {of} the fixed BS-user distance $t_{\rm bu}$. From \eqref{eq:F_Wq_UL}, the uplink compliance probability conditioned on $t_{\rm bu}$ in the direct LoS/NLoS link is either $0$ or $1$ at a given $\omega'$.  
Thus, combining the association probability with the conditional compliance probability at a fixed $t_{\rm bu}$, the overall compliance probability is shown in the blue dashed curve at the right side of Fig.~\ref{fig:marginal_UL}}.
{Moreover, compared with the network without RISs, we see that deploying RISs with $N_{\rm r}=64$ not only improves the uplink coverage performance by roughly $33\%$ at $\gamma'=-10 ~\rm dB$ but also improves the uplink compliance performance by roughly $33\%$ at $\omega'=0.3~\rm mW/kg$.} 
The reason behind the uplink coverage improvement is similar to that in the downlink case. 
The improvement in uplink compliance performance can be explained by the uplink power control mechanism in Sec.~\ref{subsec:PowerControl}, which is different from the downlink case with constant transmit power.
As indicated by \eqref{eq:PtranUL_cL}, based on the power control factor $\epsilon$, RIS-assisted uplink transmission utilizes the RIS gain to partially compensate for path loss in the cascaded LoS link. With a large enough $N_{\ rmr}$ to provide high RIS gain, the transmit power in the cascaded LoS link can be reduced, mitigating the uplink \ac{EMFE}. Fig.~\ref{fig:marginal_DL} and Fig.~\ref{fig:marginal_UL} clearly answer \textbf{Q1}, showing that deploying RISs would exacerbate downlink EMFE while alleviating uplink EMFE.

\subsubsection{Joint Probability}

Fig.~\ref{fig:UL_joint} shows the joint distribution of the uplink SINR ($\Upsilon'$) and \ac{EMFE} ($\mathcal{W}'$), i.e., $\mathcal{J}_{\Upsilon',\mathcal{W}'}(\gamma',\omega')$.
The high level of agreement between the simulation and analytical results confirms the validity of Theorem~\ref{theorem:JOINT_UL}.
We observe that for a given SINR threshold $\gamma'=-10~\rm dB$, the probability that the uplink \ac{EMFE} is below $\omega'=0.4 ~\rm mW/kg$ increases by roughly $15\%$ after deploying $16$-element RISs. In conclusion, Fig.~\ref{fig:DL_joint} and Fig.~\ref{fig:UL_joint} answer \textbf{Q3}, indicating that deploying RISs is effective in increasing both downlink and uplink SINR and mitigating uplink \ac{EMFE}, while exacerbating downlink EMFE. Moreover, deploying RISs improves joint coverage and compliance performance in both downlink and uplink scenarios.

Fig.~\ref{fig:UL_epsilon} shows the impact of the power control factor $\epsilon$ on the uplink joint performance at $\gamma'=-10\,\rm dB$ and $\omega'=0.4\,\rm mW/kg$.
With an increase in $\epsilon$, the uplink exposure (measured by the transmit power of the typical user) intensifies.
This is because the transmit power of a user is proportional to $\epsilon$, as shown in \eqref{eq:PtranUL_dL}-\eqref{eq:PtranUL_cL}. Moreover, the increased transmit power strengthens the serving signal power, while the interference from interfering users also increases. 
Under the default values of system parameters given in Table~\ref{tab:simulation}, we find a trade-off value of $\epsilon$ at $0.6$ for the RIS-assisted networks. This value offers the optimal balance between the benefits of increased serving signal power and the negative impacts of increased \ac{EMFE} and interference, which can serve as a potential uplink deployment strategy for \textbf{Q4}.
The finding emphasizes the importance of the power control factor in designing and optimizing RIS-assisted networks.

\section{Conclusions}\label{sec:conclusion}

This paper provides a framework for modeling RIS-assisted networks for the downlink and uplink and for analyzing the impact of RIS deployment on both
\ac{SINR} and \ac{EMFE} independently or jointly. 
From the numerical results, both the downlink and uplink joint distributions show that the RIS-assisted network has better performance than the network without RISs. Specifically, after deploying $16$-element RISs, the joint performance increases by roughly $15\%$ at given thresholds $\gamma=0~\rm dB$ and $\omega=0.5~\rm mW/m^2$ in downlink (or $\gamma'=-10~\rm dB$ and $\omega'=0.4~\rm mW/kg$ in uplink). These affirm the feasibility of the RIS deployment.  
However, to further enable the practical deployment, it is crucial to consider \ac{EMFE} regulations tailored to \ac{RIS}, as the marginal distribution indicates a higher downlink \ac{EMFE} level with RIS deployment.
For example, for the CD design, when the distance between a $64$-element RIS and a BS is as close as only $25~\rm m$, users should be strictly prohibited from entering the area centered at the RIS with a radius of $20 ~\rm m$ because of the excessive \ac{EMFE} within this area.
Moreover, the adjustable parameters (e.g., the densities of BSs and RISs, the number of BS antennas and RIS elements, and the power control factor) enable the performance evaluation of different configurations of RIS-assisted networks.
For example, we find a trade-off value of the power control factor, which maximizes the joint uplink performance under a specific network configuration.
Overall, our study sheds light on RIS deployment strategies to strike a balance between SINR enhancement and \ac{EMFE} management.

For future works, it would be interesting to extend the proposed framework to different types of RISs, e.g., simultaneously transmitting and reflecting (STAR)-RIS~\cite{liu2023simultaneously}.
Moreover, beam misalignment, which could potentially affect \ac{SINR} adversely and \ac{EMFE} favorably, is also a future research direction. 
Fully understanding the impact of beam misalignment caused by, e.g., the fixed codebook with finite directional beams and the imperfect channel angle information, could provide valuable insights into determining optimal codebook size and the precision required for channel estimation.

\appendices


\section{Proof of Lemma~\ref{lemma:L_SINR}}\label{app:lemma:L_SINR}
The Laplace transform of $I_{\rm B}$ conditioned on $t_{\rm bu}$ is
\begin{align}\label{eq:LIB}
\mathcal{L}_{I_{\rm B}|t_{\rm bu}} (s)&  
=\mathbb{E}\left[\exp\left(-sI_{\rm B}\right)\right]
=\mathbb{E}\left[\exp\left(-s(I_{\rm b,L}+I_{\rm b,N})\right)\right]
\nonumber\\&=\prod_{v\in\{\rm L,N\}}\mathbb{E}_{\Psi_{{\rm b},v}, H_{v,i}}\left[\exp\left(-sI_{{\rm b},v}\right)\right].
\end{align} 
Define $\mathcal{L}_{I_{{\rm b},v}|t_{\rm bu}}(s)=\mathbb{E}_{\Psi_{{\rm b},v}, H_{v,i}}\left[\exp\left(-sI_{{\rm b},v}\right)\right]$.
Then, 
\begin{align}\label{eq:L_Ibv}
&\mathcal{L}_{I_{{\rm b},v}|t_{\rm bu}} (s) 
\nonumber\\&
=\mathbb{E}_{\Psi_{{\rm b},v},H_{v,i}} \left[\exp\left(-s \!\!\! \!\!\!\sum_{i,\mathbf{b}_i\in\Psi_{{\rm b},v}\setminus\{\mathbf{b}_0\}}\!\!\! \!\!\! {p_{\rm b} G_{\rm B}(\Delta_{i})\zeta d_i^{-\alpha_{v}}H_{v,i}}
\right)\right]  
\nonumber\\&=\mathbb{E}_{\Psi_{{\rm b},v}}\left[\prod_{i,\mathbf{b}_i\in\Psi_{{\rm b},v}\setminus\{\mathbf{b}_0\}} \!\!\! \!\!\!\mathbb{E}_{H_{v,i}} [\exp\left(-s {p_{\rm b} G_{\rm B}(\Delta_{i})\zeta d_i^{-\alpha_{v}}H_{v,i}}]
\right)\right]  
\nonumber\\&\overset{(a)}{=}\mathbb{E}_{\Psi_{{\rm b},v}}\left[ \prod_{i,\mathbf{b}_i\in\Psi_{{\rm b},v}\setminus \{\mathbf{b}_0\}}
\left(\frac{m_{v}}{m_{v}+s p_{\rm b} G_{\rm B}(\Delta_{i})\zeta d_i ^{-\alpha_{v}}}\right)^{m_{v}}\right]
\nonumber\\&\overset{(b)}{=}\exp\bigg(
-2\pi\lambda_{\rm b}\int_{t_{\rm bu}}^{\infty}
[1-\hat\kappa_{v} (d_i)] d_i \mathcal{P}_{v}(d_i)
\mathrm{d} d_i
\bigg),
\end{align}
where $\hat \kappa_{v} (d_i)=\mathbb{E}_{\Delta_i}\left[\left(1+ \frac{s p_{\rm b} G_{\rm B}(\Delta_i)\zeta}{m_{v} d_i ^{\alpha_{v}}}\right)^{-m_{v}}\right]$,
(a) is from the \ac{PDF} of small-scale fading coefficient $H_{v,i}$ in \eqref{eq:ssfading}, and (b) is from the probability generating functional (PGFL) of the \ac{PPP}~\cite{andrews2016primer} with the density of $\Psi_{{\rm b},v}$ being $\lambda_{\rm b}\mathcal{P}_{v}(d_i)$.  
Based on $f_{\Delta}(\cdot)$ in \eqref{eq:pdf_Delta}, we can further express $\hat \kappa_{v} (d_i)$ as
\begin{align}\label{eq:kappa}
    \begin{split}
        \hat \kappa_{v} (d_i)&=\int_{-1}^{1}
\!\!\left(1+ \frac{s p_{\rm b} G_{\rm B}(\Delta_i)\zeta}{m_{v} d_i ^{\alpha_{v}}}\right)^{-m_{v}} f_{\Delta}(\Delta_i)\mathrm{d} \Delta_i
\\& \overset{(a)}{=}\sum_{n=0}^{\tilde{N}_{\rm b}+1}
p_n\left(1+ \frac{s p_{\rm b} G_n \zeta}{m_{v} d_i ^{\alpha_{v}}}\right)^{-m_{v}},
    \end{split}
\end{align}
where $p_n$ is given in \eqref{eq:pn} and (a) is from Lemma~\ref{lemma:prob_Gain}.
Substituting \eqref{eq:kappa} and \eqref{eq:L_Ibv} into \eqref{eq:LIB}, we finish the proof.


\section{Proof of Theorem~\ref{theorem:emf}}\label{app:theorem:emf}
The \ac{CDF} of the overall \ac{EMFE} $\mathcal{W}$ can be derived based on the conditional compliance probability in \eqref{eq:F_Wq}.
The first moment of $\mathcal{W}$ is defined as
 \begin{align} \label{eq:E_W0}
\begin{split}
& \mathbb{E}[\mathcal{W}]  = \mathbb{E}\bigg[ \sum_{q\in\mathcal{S}} {\mathcal{W}_{q}} \mathcal{A}_{q}\bigg] 
\overset{(a)}{=}  \mathbb{E}\bigg[ \sum_{q\in\mathcal{S}} \mathcal{W}_{1,q}\mathcal{A}_{q}+\mathcal{W}_{2} \bigg]
\\& \overset{(b)}{=}  \mathbb{E}\bigg[ \sum_{q\in\mathcal{S}}  \mathcal{W}_{1,q}\mathcal{A}_{q} +\mathcal{W}_{\rm b,L} + \mathcal{W}_{\rm b,N}\bigg] 
\\&= \mathbb{E}_{t_{\rm bu},t_{\rm ru},t_{\rm br}}\bigg[ \sum_{q\in\mathcal{S}} \mathcal{A}_{q} \mathbb{E}_{H_q} [\mathcal{W}_{1,q}]  \bigg] 
+ \\& ~\quad \mathbb{E}_{t_{\rm bu}}\bigg[ \mathbb{E}_{\Psi_{{\rm b,L}},\Delta_{i},H_{{\rm L},i}}[\mathcal{W}_{\rm b,L}]+\mathbb{E}_{\Psi_{{\rm b,N}},\Delta_{i},H_{{\rm N},i}}[\mathcal{W}_{\rm b,N}] \bigg],
\end{split}
\end{align}
where (a) is from \eqref{eq:Wq}, and (b) is from \eqref{eq:W2}.
With \eqref{eq:Wb0} and $\mathbb{E}[H_q]=1$, we have 
\begin{align} \label{eq:E_W1q}
    \mathbb{E}_{H_q} [\mathcal{W}_{1,q}]=\mathbb{E}_{H_q} \bigg[{P^{\rm a}_{q} H_q}{\mathcal{E}^{-1}}\bigg] = {P^{\rm a}_{q}}{\mathcal{E}^{-1}} . 
\end{align}
From \eqref{eq:W2}, \eqref{eq:Ibv}, and ${\mathcal{E}}/{\zeta}=4\pi$, we have
\begin{align} \label{eq:E_WbL}
&\mathbb{E}_{\Psi_{{\rm b,L}},\Delta_{i},H_{{\rm L},i}}[\mathcal{W}_{\rm b,L}] \nonumber
\\& =
\mathbb{E}_{\Psi_{{\rm b,L}},\Delta_{i},H_{{\rm L},i}}\bigg[\sum_{i,\mathbf{b}_i\in\Psi_{{\rm b,L}}\setminus\{\mathbf{b}_0\}}
  \!\!\!\!\frac{p_{\rm b} G_{\rm B}(\Delta_{i})H_{{\rm L},i}}{4\pi d_i ^{\alpha_{\rm L}}}\bigg]
\nonumber\\&\overset{(a)}{=}\int_{t_{\rm bu}}^{\infty}\mathbb{E}_{\Delta_{i},H_{{\rm L},i}}\bigg[
 \frac{p_{\rm b} G_{\rm B}(\Delta_{i})H_{{\rm L},i}}{4\pi d_i ^{\alpha_{\rm L}}}\bigg]  2\pi \lambda_{\rm b} \mathcal{P}_{\rm L}(d_i)d_i \mathrm{d} d_i
\nonumber\\&=\int_{t_{\rm bu}}^{\infty}  
 \frac{p_{\rm b} \mathbb{E}_{\Delta} [G_{\rm B}(\Delta)] \mathbb{E}[H_{{\rm L}}]}{2 d_i ^{\alpha_{\rm L}}} \lambda_{\rm b} \mathcal{P}_{\rm L}(d_i)d_i \mathrm{d} d_i
\nonumber\\&\overset{(b)}{=}  \frac{\lambda_{\rm b}  p_{\rm b} \bar g_{\rm B} }{2} \int_{t_{\rm bu}}^{\infty}  
 \frac{\exp\left(-\beta t \right) }{t ^{\alpha_{\rm L}-1}} \mathrm{d} t,
\end{align}
where (a) applies Campbell Theorem~\cite{SGtutorial} with the density of $\Psi_{{\rm b,L}}$ being $\lambda_{\rm b}\mathcal{P}_{\rm L}(d_i)$, (b) replaces $d_{i}$ with $t$, (b) is also from $\mathcal{P}_{\rm L}(t)$ in \eqref{eq:PLoS}, $\mathbb{E}[H_{{\rm L}}]=1$, and $\bar g_{\rm B} = \mathbb{E}_{\Delta} [G_{\rm B}(\Delta)] =\sum_{n=0}^{\tilde{N}_{\rm b}} G_np_n$ is from Lemma~\ref{lemma:prob_Gain}. 
Similarly, considering that the density of $\Psi_{{\rm b,N}}$ is $\lambda_{\rm b}\mathcal{P}_{\rm N}(d_i)=\lambda_{\rm b}(1-\mathcal{P}_{\rm L}(d_i))$, we have
 \begin{align} \label{eq:E_WbN}
&\mathbb{E}_{\Psi_{{\rm b,N}},\Delta_{i},H_{{\rm N},i}}[\mathcal{W}_{\rm b,N}]
=  \frac{\lambda_{\rm b}  p_{\rm b} \bar g_{\rm B} }{2} \!\!\int_{t_{\rm bu}}^{\infty}  
 \frac{1-\exp\left(-\beta t \right) }{t ^{\alpha_{\rm N}-1}} \mathrm{d} t.
\end{align}
Note that $\int_{a}^{\infty}  
 \frac{\exp\left(-\beta t \right) }{t ^{n} } \mathrm{d} t 
= a^{1-n} {\rm E}_n(\beta a)$, and ${\rm E}_n(x) = \int_{1}^{\infty}  
 \frac{\exp\left(-x t \right) }{t ^{n} } \mathrm{d} t $. 
Substituting \eqref{eq:E_W1q}-\eqref{eq:E_WbN} into \eqref{eq:E_W0}, we prove \eqref{eq:E_W}. 
\section{Proof of Proposition~\ref{prop:CD}}\label{app:CD}
From \eqref{eq:Wq} and \eqref{eq:F_Wq}, considering the dominant downlink \ac{EMFE} from the serving  link,  
we approximate the BS CD in \eqref{eq:xcom} as
\begin{align}\label{eq:xcom1}
\begin{split}
\tau_{\rm bu}&=\inf_{t_{\rm bu}\in \mathbb{R}} \left\{t_{\rm bu}:   \mathbb{P}(\mathcal{W}_{1,\rm DL}+\mathcal{W}_{2}\le \mathcal{W}_{\max} | t_{\rm bu})  \geq \rho  \right\}
\\& \approx  \inf_{t_{\rm bu}\in \mathbb{R}} \left\{t_{\rm bu}:   \mathbb{P}(\mathcal{W}_{1,\rm DL} \le \mathcal{W}_{\max} | t_{\rm bu})  \geq \rho  \right\}
\\& \triangleq  \inf_{t_{\rm bu}\in \mathbb{R}} \left\{t_{\rm bu}:   F_{\mathcal{W}_{1,\rm DL}|t_{\rm bu}}(\mathcal{W}_{\max}) \geq \rho  \right\}.
\end{split}
\end{align}
From \eqref{eq:Wb0}, we have 
\begin{align} \label{eq:F_W1DL}\nonumber
&F_{\mathcal{W}_{1,\rm DL}|t_{\rm bu}} (\omega) = \mathbb{P}(\mathcal{W}_{1,\rm DL}\leq \omega|t_{\rm bu}) 
= \mathbb{P}\bigg( \frac{P^{\rm a}_{\rm DL} H_{\rm DL}}{\mathcal{E}} \leq \omega\bigg|t_{\rm bu}\bigg)  
\\&= \mathbb{P}\bigg( H_{\rm DL}  \leq \frac{\mathcal{E} \omega}{P^{\rm a}_{\rm DL} }\bigg|t_{\rm bu}\bigg) 
=F_{H_{\rm L}}\bigg(\frac{\mathcal{E} \omega}{P^{\rm a}_{\rm DL} }\bigg),
\end{align}
where $F_{H_{\rm L}}(\cdot)$ is the PDF of $H_{\rm DL}$ given in \eqref{eq:CDF_ssfading} and $P^{\rm a}_{\rm DL} $ is given in \eqref{eq:Pr_dLN}.
With \eqref{eq:F_W1DL} and \eqref{eq:Pr_dLN}, \eqref{eq:xcom1} can be expressed as
\begin{align}
    \tau_{\rm bu} &\approx  \inf_{t_{\rm bu}\in \mathbb{R}} \left\{t_{\rm bu}:  F_{H_{\rm L}}\bigg(\frac{\mathcal{E} \mathcal{W}_{\max}}{ p_{\rm b} G_{\rm b} \zeta t_{\rm bu}^{-\alpha_{\rm L}} }\bigg)  \geq \rho  \right\}
\nonumber\\&= \bigg(\frac{ p_{\rm b} G_{\rm b} \zeta}{\mathcal{E} \mathcal{W}_{\max}} F_{H_{\rm L}}^{-1}(\rho)\bigg)^{\frac{1}{\alpha_{\rm L}}} 
\!\!\!\!= \!\!\bigg(\frac{ p_{\rm b} G_{\rm b} }{ 4\pi \mathcal{W}_{\max}} F_{H_{\rm L}}^{-1}(\rho)\bigg)^{\frac{1}{\alpha_{\rm L}}} .  
\end{align}
Similar to \eqref{eq:F_W1DL}, we define the CDF of the \ac{EMFE} from the cascaded LoS link as 
\begin{align} \label{eq:F_W1CL}\nonumber
F_{\mathcal{W}_{1,\rm CL}|t_{\rm ru},t_{\rm br}} (\omega) &= \mathbb{P}(\mathcal{W}_{1,\rm CL}\leq \omega|t_{\rm ru},t_{\rm br}) 
=F_{H_{\rm L}}\bigg(\frac{\mathcal{E} \omega}{P^{\rm a}_{\rm CL} }\bigg),
\end{align}
where $P^{\rm a}_{\rm CL}$ is given in \eqref{eq:Pr_cL}.
Then, we approximate the conditional RIS CD in \eqref{eq:ycom} as
\begin{align}
 \hat \tau_{\rm ru}(t_{\rm br}) 
& \approx \inf_{t_{\rm ru}\in \mathbb{R}} \left\{t_{\rm ru}:   F_{\mathcal{W}_{1,\rm CL}|t_{\rm ru},t_{\rm br}}(\mathcal{W}_{\max}) \geq \rho  \right\} 
\nonumber\\&= \bigg(\frac{ p_{\rm b} G_{\rm b} G_{\rm r} t_{\rm br}^{-\alpha_{\rm L}} } {4\pi \mathcal{W}_{\max}} F_{H_{\rm L}}^{-1}( \rho)
\bigg)^{\frac{1}{\alpha_{\rm L}}}.
\end{align}
Similarly, for the average RIS CD, 
\begin{align}\label{eq:RIS_CD_aver0}
\begin{split}
    \tau_{\rm ru}  &\approx \inf_{t_{\rm ru}\in \mathbb{R}} \left\{t_{\rm ru}:   \mathbb{E}_{t_{\rm br}} [ F_{\mathcal{W}_{1,\rm CL}|t_{\rm ru},t_{\rm br}}(\mathcal{W}_{\max}) ] \geq \rho   \right\}.
\end{split}
\end{align}
The approximation in \eqref{eq:RIS_CD_aver0} implies that $\tau_{\rm ru} $ satisfies
\begin{align}\label{eq:RIS_CD_aver1}
    \mathbb{E}_{t_{\rm br}} \bigg[ F_{H_{\rm L}}\bigg( \frac{4\pi\mathcal{W}_{\max}}{ p_{\rm b} G_{\rm b} G_{\rm r}  (\tau_{\rm ru}t_{\rm br})^{-\alpha_{\rm L}} } \bigg) \bigg]=\rho.
\end{align}
In \eqref{eq:RIS_CD_aver0} and \eqref{eq:RIS_CD_aver1}, 
given $t_{\rm ru}$,  we observe from \eqref{eq:z0} that  $t_{\rm br}$ is a random variable whose distribution depends on $t_{\rm bu}$ and $\theta_0$. This greatly complicates the calculation of $\mathbb{E}_{t_{\rm br}} [ F_{H_{\rm L}}( h_0 t_{\rm br}^{\alpha_{\rm L}} ) ]$, where 
\begin{align}\label{eq:h0}
h_0= {4 \pi \mathcal{W}_{\max}}{ (p_{\rm b} G_{\rm b} G_{\rm r} \tau_{\rm ru}^{-\alpha_{\rm L}})^{-1} }.
\end{align}
To simplify the calculation of $\mathbb{E}_{t_{\rm br}} [\cdot]$ , we notice that $t_{\rm br}$ satisfies the triangle inequality $|t_{\rm bu}-t_{\rm ru}| \le t_{\rm br} \le t_{\rm bu}+t_{\rm ru}$. Considering that the value of $\tau_{\rm ru}$ is generally small, when $t_{\rm ru}=\tau_{\rm ru}$, we can expect that $t_{\rm br} \approx t_{\rm bu}$. Inspired by this, we use the PDF of $t_{\rm bu}$, i.e., $f_{T_{\rm bu}}(\cdot)$ in \eqref{eq:fX}, to approximate the distribution of $t_{\rm br}$.  Then, $\mathbb{E}_{t_{\rm br}} [ F_{H_{\rm L}}( h_0 t_{\rm br}^{\alpha_{\rm L}} ) ]$ in \eqref{eq:RIS_CD_aver1} becomes
\begin{align}\label{eq:RIS_CD_aver2}
&\mathbb{E}_{t_{\rm br}} [ F_{H_{\rm L}}( h_0 t_{\rm br}^{\alpha_{\rm L}} ) ]
\overset{(a)}{\approx} \int_{0}^{\infty} F_{H_{\rm L}}( h_0 t_{\rm br}^{\alpha_{\rm L}}  ) f_{T_{\rm bu}}(t_{\rm br}) \mathrm{d} t_{\rm br} 
\nonumber\\& \overset{(b)}{=}  \int_{0}^{\infty} 
\frac{\Gamma_l(m_{\rm L},m_{\rm L}h_0 t^{\alpha_{\rm L}}) }{\Gamma(m_{\rm L})} 2\pi\lambda_{\rm b}t\exp(-\pi\lambda_{\rm b}t^2 ) \mathrm{d} t 
\nonumber\\& \overset{(c)}{=} -
\frac{\Gamma_l(m_{\rm L},m_{\rm L}h_0 t^{\alpha_{\rm L}}) }{\Gamma(m_{\rm L})} \exp(-\pi\lambda_{\rm b}t^2 ) \bigg |_{0}^{\infty} +
\nonumber\\& \qquad \int_{0}^{\infty}  \exp(-\pi\lambda_{\rm b}t^2 )   \mathrm{d} \frac{\Gamma_l(m_{\rm L},m_{\rm L}h_0 t^{\alpha_{\rm L}}) }{\Gamma(m_{\rm L})} 
\nonumber\\& \overset{(d)}{=}  \frac{\alpha_{\rm L}(m_{\rm L}h_0)^{m_{\rm L}}}{\Gamma(m_{\rm L})} \!\!\int_{0}^{\infty} 
\!\! t^{m_{\rm L}\alpha_{\rm L}-1} e^{-m_{\rm L}h_0 t^{\alpha_{\rm L}} -\pi\lambda_{\rm b}t^2 }  \mathrm{d} t  
\nonumber\\& \triangleq F_{\alpha_{\rm L}}(h_0) ,
\end{align}
where (a) is from  $t_{\rm br}\approx t_{\rm bu}$ when $t_{\rm ru}=\tau_{\rm ru}$ is small, (b) replaces $t_{\rm br}$ with $t$, (b) is also from \eqref{eq:CDF_ssfading} and \eqref{eq:fX}, (c) is from integration by parts, and (d) is from 
$\mathrm{d} \Gamma_l(m,x) = \mathrm{d} \int_{0}^{x} z^{m-1} e^{-z} \mathrm{d} z = x^{m-1} e^{-x}$. 
From \eqref{eq:RIS_CD_aver2} and \eqref{eq:RIS_CD_aver1}, 
we have $h_0= F_{\alpha_{\rm L}}^{-1}(\rho)$, where $F_{\alpha_{\rm L}}^{-1}(\rho)$ is the inverse function of $F_{\alpha_{\rm L}}(h_0)$.
Then, with \eqref{eq:h0}, we obtain \eqref{eq:CDru}.
Furthermore, when $\alpha_{\rm L}=2$, we obtain
\begin{align}\label{eq:alpha=2}
F_{\alpha_{\rm L}=2}(h_0) 
&= \frac{2(m_{\rm L}h_0)^{m_{\rm L}} }{\Gamma(m_{\rm L})} \int_{0}^{\infty}  
t^{2m_{\rm L}-1} e^{-(m_{\rm L}h_0+\pi\lambda_{\rm b}) t^{2}} 
\mathrm{d} t
\nonumber\\& \overset{(a)}{=} \frac{(m_{\rm L}h_0)^{m_{\rm L}} }{\Gamma(m_{\rm L})} \int_{0}^{\infty}
z^{m_{\rm L}-1} e^{-(m_{\rm L}h_0+\pi\lambda_{\rm b}) z} 
\mathrm{d} z
\nonumber\\& \overset{(b)}{=} \frac{(m_{\rm L}h_0)^{m_{\rm L}} }{\Gamma(m_{\rm L}) (m_{\rm L}h_0+\pi\lambda_{\rm b})^{m_{\rm L}}} \int_{0}^{\infty}
x^{m_{\rm L}-1} e^{-x} 
\mathrm{d} x
\nonumber\\& \overset{(c)}{=}  {(m_{\rm L}h_0)^{m_{\rm L}} }{(m_{\rm L}h_0+\pi\lambda_{\rm b})^{-m_{\rm L}}} ,
\end{align}
where (a) replaces $t^2$ with $z$, (b) replaces $(m_{\rm L}h_0+\pi\lambda_{\rm b}) z$ with $x$, and (c) is from $\Gamma(m)=\int_{0}^{\infty} x^{m-1} e^{-x} \mathrm{d}x$.
Then, $F_{\alpha_{\rm L}=2}(h_0) = \rho$ implies that
\begin{align}\label{eq:h1}
\begin{split}
h_0=\frac{\pi\lambda_{\rm b} \rho^{\frac{1}{m_{\rm L}}} }{m_{\rm L}\big(1-\rho^{\frac{1}{m_{\rm L}}}\big)} ~{\rm for }~ \alpha_{\rm L}=2.
\end{split}
\end{align}
Substituting \eqref{eq:h1} into \eqref{eq:h0}, we obtain \eqref{eq:CDru_alpha2}.

\section{Proof of Theorem~\ref{theorem:Joint}}\label{app:lemma:JOINT_q}
{Certain techniques for deriving the joint metric have been explored in \cite{R1-4,R1-5,gontier2023joint}, which can be 
adapted for deriving Theorem~\ref{theorem:Joint} in our work.}
Specifically, when the serving link is direct LoS, the downlink joint probability conditioned on $t_{\rm bu}$ is 
\begin{align}\label{eq:joint_1}
&J_{\Upsilon_{\rm DL},\mathcal{W}_{\rm DL}|t_{\rm bu}}(\gamma,\omega)=\mathbb{P}({\Upsilon_{\rm DL}}>\gamma,{\rm \mathcal{W}_{\rm DL}}\le \omega |t_{\rm bu})
\nonumber\\&\overset{(a)}{=}\mathbb{P}\left(\frac{P_{\rm DL}}{\sigma^2+I_{\rm B}}>\gamma,\frac{P_{\rm DL}+I_{\rm B}}{\mathcal{E}}\le \omega\right)
\nonumber\\&\overset{(b)}{=}\mathbbm{1}(\mathcal{E} \omega - \sigma^2 \gamma  \ge 0)\mathbb{P}\left(I_{\rm B}<\frac{P_{\rm DL}}{\gamma}-\sigma^2,I_{\rm B}\le \mathcal{E} \omega -P_{\rm DL}\right)
\nonumber\\&\overset{(c)}{=}\mathbbm{1}(\mathcal{E} \omega - \sigma^2 \gamma  \ge 0)\times
\nonumber\\&\quad\begin{cases}
 \mathbb{P}(I_{\rm B}<\frac{P_{\rm DL}}{\gamma}-\sigma^2),
& \text{ if }  \sigma^2 \gamma  \le P_{\rm DL} < P_1,\\
 \mathbb{P}(I_{\rm B}\le \mathcal{E} \omega -P_{\rm DL}),
& \text{ if } P_1\le P_{\rm DL} \le\mathcal{E} \omega,\\
0,
& \text{ otherwise,}  
\end{cases}
\end{align} 
where $P_1=\frac{\gamma (\mathcal{E} \omega+\sigma^2)}{1+\gamma}$, (a) is from \eqref{eq:Wq}, and (b) and (c) is from the non-negativity of interference (i.e., $I_{\rm B}\ge 0$)~\cite{gontier2023joint}. Specifically, either $P_{\rm DL} < \sigma^2 \gamma $ or $P_{\rm DL}> \mathcal{E} \omega $ leads to $I_{\rm B}<0$.  
Therefore, when $\mathcal{E} \omega < \sigma^2 \gamma $,  $J_{\Upsilon_{\rm DL},\mathcal{W}_{\rm DL}|t_{\rm bu}}(\gamma,\omega)=0$. {Moreover, when $\mathcal{E} \omega \ge \sigma^2 \gamma $,  $\sigma^2 \gamma  \le P_1 \le \mathcal{E} \omega$.}
Let $U(\gamma,\omega)=\mathbbm{1}(\mathcal{E} \omega - \sigma^2 \gamma \ge 0)$. 
Substituting \eqref{eq:Pr_dLN} into \eqref{eq:joint_1}, we have 
 \begin{align}\label{eq:joint_2}
&J_{\Upsilon_{\rm DL},\mathcal{W}_{\rm DL}|t_{\rm bu}}(\gamma,\omega)= U(\gamma,\omega) \times 
\nonumber\\&
\Bigg(\mathbb{E}\left[\mathbbm{1}\left(\frac{\sigma^2 \gamma }{P^{\rm a}_{\rm DL}} \le  H_{\rm DL} <\frac{P_1}{P^{\rm a}_{\rm DL}}\right)
\mathbb{P}\left(I_{\rm B}<\frac{P^{\rm a}_{\rm DL} H_{\rm DL}}{\gamma}-\sigma^2\right)
\right]
\nonumber\\&
~+\mathbb{E}\left[\mathbbm{1}\left(\frac{P_1}{P^{\rm a}_{\rm DL}} \le  H_{\rm DL} \le\frac{\mathcal{E} \omega}{P^{\rm a}_{\rm DL}}\right)
\mathbb{P}(I_{\rm B}< \mathcal{E} \omega  - P^{\rm a}_{\rm DL}H_{\rm DL})
\right] \Bigg)
\nonumber\\&=U(\gamma,\omega) \times 
\Bigg(\underbrace{ \int_{\frac{\sigma^2 \gamma }{P^{\rm a}_{\rm DL}} }^{\frac{P_1}{P^{\rm a}_{\rm DL}}}
F_{I_{\rm B}|t_{\rm bu}} \bigg(\frac{P^{\rm a}_{\rm DL}h}{\gamma}-\sigma^2\bigg) f_{H_{\rm L}}(h) \mathrm{d}h
}_{\mathcal{J}^{\rm DL}_1}
\nonumber\\& ~+ 
\underbrace{ \int^{\frac{\mathcal{E} \omega}{P^{\rm a}_{\rm DL}}}_{\frac{P_1}{P^{\rm a}_{\rm DL}}}
F_{I_{\rm B}|t_{\rm bu}} \bigg( \mathcal{E} \omega  - P^{\rm a}_{\rm DL} h\bigg) f_{H_{\rm L}}(h) \mathrm{d}h
}_{\mathcal{J}^{\rm DL}_2} \Bigg),
\end{align}  
where $F_{I_{\rm B}|t_{\rm bu}}(p)=\mathbb{P}(I_{\rm B}\leq p|t_{\rm bu})$ is the CDF of $I_{\rm B}$ conditioned on $t_{\rm bu}$ and $f_{H_{\rm L}}(h)$ is the PDF of $H_{\rm DL}$ {given in \eqref{eq:ssfading}.}
It is worth noting that for a continuous variable $H$, the CDF of $H$ is $F_H(h)=\mathbb{P}(H \le h)=\mathbb{P}(H < h)$. 
{Therefore, the replacement between ``$<$'' and ``$\le$'' in \eqref{eq:joint_2} does not affect the results.}
In the following, we derive $F_{I_{\rm B}|t_{\rm bu}}(p)$ and $f_{H_{\rm L}}(h)$ to obtain the final expression of \eqref{eq:joint_2}.
From Gil-Pelaez theorem, the CDF of $I_{\rm B}$ is 
\begin{equation}\label{eq:CDF_IB}
\begin{split}
F_{I_{\rm B}|t_{\rm bu}}(p)&=
\frac{1}{2} -\int_{0}^{\infty}\frac{1}{ \pi x}{\rm Im}\left [ e^{-jx p}\mathcal{L}_{I_{\rm B}|t_{\rm bu}}(-jx) \right ] {\rm d} x, 
\end{split}
\end{equation}
where $\mathcal{L}_{I_{\rm B}|t_{\rm bu}}(\cdot)$ is given in \eqref{eq:LIB}.
By noting that shaping parameter $m_{\rm L}$ is an integer, the CDF of $H_{\rm DL}$ in \eqref{eq:ssfading} can be further expressed as 
\begin{align}\label{eq:cdf_HL}
F_{H_{\rm L}}(h) 
\overset{(a)}{=}
1-\frac{\Gamma_u(m_{\rm L},m_{\rm L}h)}{\Gamma(m_{\rm L})}
\overset{(b)}{=}1-e^{-m_{\rm L}h}\sum_{k=0}^{m_{\rm L}-1}\frac{(m_{\rm L}h)^k}{k!},
\end{align}
where {$\Gamma_u\left ( m,mg \right )=\int_{mg}^{\infty} x^{m-1}e^{-x}\mathrm{d}x$,}
(a) is from $\Gamma_l\left ( m,mg \right )=\Gamma\left ( m\right )-\Gamma_u\left ( m,mg \right )$, and (b) is from the fact that $\frac{\Gamma_u\left ( m,g \right )}{\Gamma\left ( m \right )} =\exp(-g) {\textstyle \sum_{k=0}^{m-1}}\frac{g^k}{k!}$ for $m\in \mathbb{N}$.
Then, we can obtain the PDF of $H_{\rm DL}$ as $f_{H_{\rm L}}(h)  =\frac{\mathrm{d} F_{H_{\rm L}}(h)}{\mathrm{d} h}$, i.e., 
\begin{align}\label{eq:pdf_HL}
&f_{H_{\rm L}}(h)  
\!=\!-\!\!\!\sum_{k=0}^{m_{\rm L}-1}\!\frac{m_{\rm L}^k}{k!} \left(
-m_{\rm L}e^{-m_{\rm L}h} h^k+ k e^{-m_{\rm L}h} h^{k-1}
\right).\!
\end{align} 
Substituting \eqref{eq:CDF_IB}-\eqref{eq:pdf_HL} into $\mathcal{J}^{\rm DL}_1$, we have 
\begin{align}\label{eq:JdL1}
&\mathcal{J}^{\rm DL}_1 
=\frac{1}{2} F_{H_{\rm L}}\bigg( \frac{P_1}{ P^{\rm a}_{\rm DL} } \bigg) - \frac{1}{2} F_{H_{\rm L}}\bigg( \frac{\sigma^2 \gamma }{ P^{\rm a}_{\rm DL} } \bigg) -
\nonumber\\&\int_{0}^{\infty}\frac{1}{\pi x}{\rm Im}\left [ 
\Xi^{\rm DL}_1(-jx)
\exp(jx\sigma^2)\mathcal{L}_{I_{\rm B}|t_{\rm bu}}(-jx) \right ] {\rm d} x , 
\end{align}
where $\Xi^{\rm DL}_1(-jx)=\int_{\frac{\sigma^2 \gamma }{ P^{\rm a}_{\rm DL} } }^{\frac{P_1}{ P^{\rm a}_{\rm DL} }}\exp\left(-jx \frac{ P^{\rm a}_{\rm DL} h}{\gamma} \right) f_{H_{\rm L}}(h) \mathrm{d}h$.
Substituting \eqref{eq:pdf_HL} into $\Xi^{\rm DL}_1(-jx)$, we have 
\begin{align}\label{eq:Xi1_1}
&\Xi^{\rm DL}_1(-jx)  = 
\exp\left(\!-jx \frac{P^{\rm a}_{\rm DL} h}{\gamma} \right) (F_{H_{\rm L}}(h)-1) \bigg|_{\frac{\sigma^2 \gamma }{P^{\rm a}_{\rm DL}} }^{\frac{P_1}{P^{\rm a}_{\rm DL}}}
\!-jx \frac{P^{\rm a}_{\rm DL} }{\gamma} \times
\nonumber \\&\quad
\sum_{k=0}^{m_{\rm L}-1}\frac{m_{\rm L}^k}{k!} 
\int_{\frac{\sigma^2 \gamma }{P^{\rm a}_{\rm DL}} }^{\frac{P_1}{P^{\rm a}_{\rm DL}}} 
 h^k\exp\left(-\frac{jx P^{\rm a}_{\rm DL}+m_{\rm L}\gamma}{\gamma} h \right)  \mathrm{d} h.
\end{align}
Note that 
\begin{align}\label{eq:int}
\int_{a_1}^{a_2} \!\!h^k \exp(-bh) \mathrm{d} h = -\frac{\exp(-bh)}{k+1} 
\sum_{\iota =0}^{k}\frac{h^{k-\iota }}{b^{\iota +1}} A^{\iota +1}_{k+1} \bigg|_{a_1}^{a_2}, 
\end{align}
where $A^{\iota+1}_{k+1}=\frac{(k+1)!}{(k-\iota )!}$. Base on \eqref{eq:int}, we can further express \eqref{eq:Xi1_1} as \eqref{eq:xi1}. 
Similarly, 
$\mathcal{J}^{\rm DL}_2
=\frac{1}{2} F_{H_{\rm L}} \big( \frac{P_1}{ P^{\rm a}_{{\rm DL}} } \big) -  \frac{1}{2} F_{H_{\rm L}}\big( \frac{\mathcal{E} \omega}{P^{\rm a}_{{\rm DL}}} \big)  -
\int_{0}^{\infty}\frac{1}{\pi x}{\rm Im}\left [ 
\Xi^{\rm DL}_2(jx)
\exp(-jx \mathcal{E} \omega )\mathcal{L}_{I_{\rm B}|t_{\rm bu}}(-jx) \right ] {\rm d} x$, 
where $\Xi^{\rm DL}_2(jx)= \int^{\frac{\mathcal{E} \omega}{P^{\rm a}_{{\rm DL}}} }_{\frac{P_1}{P^{\rm a}_{{\rm DL}}}} \exp\left(jx P^{\rm a}_{{\rm DL}} h\right) f_{H_{\rm L}}(h) \mathrm{d}h$ can be calculated as \eqref{eq:xi2}  by using the same methods in the calculation of $\Xi^{{\rm DL}}_1(-jx)$.
Substituting $\mathcal{J}^{\rm DL}_1$ and $\mathcal{J}^{\rm DL}_2$ into \eqref{eq:joint_2}, 
we finish the derivation of $J_{\Upsilon_{\rm DL},\mathcal{W}_{\rm DL}|t_{\rm bu}}(\gamma,\omega)$.
Following the above steps, we can derive $J_{\Upsilon_{\rm DN},\mathcal{W}_{\rm DN}|t_{\rm bu}}(\gamma,\omega)$ and $J_{\Upsilon_{\rm CL},\mathcal{W}_{\rm CL}|t_{\rm bu},t_{\rm ru},t_{\rm br}}(\gamma,\omega)$. 
With \eqref{eq:joint}, we complete the proof of Theorem~\ref{theorem:Joint}.

\bibliographystyle{IEEEtran}
\bibliography{reference}

\end{document}